\documentclass[review]{elsarticle}

\usepackage{amssymb}

\usepackage{chngcntr}
\usepackage{color}
\usepackage{graphicx}
\usepackage{dcolumn}
\usepackage{bm}
\usepackage{mathrsfs}
\usepackage[ruled,vlined]{algorithm2e}
\usepackage{array} %
\usepackage[utf8]{inputenc}
\usepackage{tabstackengine}
\usepackage{amsmath}
\usepackage{amsfonts}
\setstackEOL{\cr}
\usepackage{amsfonts,amsmath,mathrsfs,booktabs}
\usepackage{multirow,times}
\usepackage{float,color,bm,subfigure}
\usepackage{empheq}
\usepackage{hhline}
\usepackage{tikz}
\usepackage{lipsum,mathtools}
\usepackage{bbm}
\usepackage[colorlinks,
linkcolor=blue,
anchorcolor=blue,
urlcolor=blue,
citecolor=blue
]{hyperref}

\newcommand{\splitatcommas}[1]{%
	\begingroup
	\ifnum\mathcode`,="8000
	\else
	\begingroup\lccode`~=`, \lowercase{\endgroup
		\edef~{\mathchar\the\mathcode`, \penalty0 \noexpand\hspace{0pt plus 0.3em}}%
	}\mathcode`,="8000
	\fi
	#1%
	\endgroup
}
\newcommand{\tuple}[1]{(\splitatcommas{#1})}
\newcommand{\set}[1]{\{\splitatcommas{#1}\}}
\newcommand{\RNum}[1]{\uppercase\expandafter{\romannumeral #1\relax}} 

\newcommand{\PreserveBackslash}[1]{\let\temp=\\#1\let\\=\temp}
\newcolumntype{C}[1]{>{\PreserveBackslash\centering}p{#1}}
\newcolumntype{R}[1]{>{\PreserveBackslash\raggedleft}p{#1}}
\newcolumntype{L}[1]{>{\PreserveBackslash\raggedright}p{#1}}
 \normalsize

\definecolor{domain_c}{rgb}{1, 0.647, 0}
\definecolor{domain_d}{rgb}{1, 0, 0}
\definecolor{wall_c}{rgb}{1, 1, 1}
\definecolor{wall_d}{rgb}{0.5, 0.5, 0.5}
\definecolor{disorder_c}{rgb}{0, 1, 0}
\definecolor{disorder_d}{rgb}{0, 0.5, 0}
\DeclareRobustCommand\domainc{\raisebox{.5ex}{\begin{tikzpicture}[baseline=(char.base)]
			\node[fill=domain_c, rectangle, inner sep=4pt] (char) {};
\end{tikzpicture}}}
\DeclareRobustCommand\domaind{\raisebox{.5ex}{\begin{tikzpicture}[baseline=(char.base)]
			\node[fill=domain_d, rectangle, inner sep=4pt] (char) {};
\end{tikzpicture}}}
\DeclareRobustCommand\wallc{\raisebox{.5ex}{\begin{tikzpicture}[baseline=(char.base)]
			\node[draw=black, fill=wall_c, rectangle, inner sep=4pt] (char) {};
\end{tikzpicture}}}
\DeclareRobustCommand\walld{\raisebox{.5ex}{\begin{tikzpicture}[baseline=(char.base)]
			\node[fill=wall_d, rectangle, inner sep=4pt] (char) {};
\end{tikzpicture}}}
\DeclareRobustCommand\disorderc{\raisebox{.5ex}{\begin{tikzpicture}[baseline=(char.base)]
			\node[fill=disorder_c, rectangle, inner sep=4pt] (char) {};
\end{tikzpicture}}}
\DeclareRobustCommand\disorderd{\raisebox{.5ex}{\begin{tikzpicture}[baseline=(char.base)]
			\node[fill=disorder_d, rectangle, inner sep=4pt] (char) {};
\end{tikzpicture}}}

\journal{Chaos, Solitons $\&$ Fractals}

\begin{document}

\begin{frontmatter}



\title{Emergence of anti-coordinated patterns in snowdrift game by reinforcement learning}




\author[1]{Zhen-Wei Ding\fnref{fn1}}
\author[1]{Ji-Qiang Zhang\fnref{fn1}\corref{cor1}}
\ead{zhangjq13@lzu.edu.cn}
\author[2]{Guo-Zhong Zheng}
\author[3]{Wei-Ran Cai}s
\author[4]{Chao-Ran Cai}
\author[2]{Li Chen}
\author[1]{Xu-Ming Wang}
\cortext[cor1]{Corresponding author.}
\fntext[fn1]{Co-first authors.}
\address[1]{School of Physics, Ningxia University, Yinchuan 750021, P. R. China}
\address[2]{School of Physics and Information Technology, Shaanxi Normal University, Xi'an, 710062, P. R. China}
\address[3]{School of Computer Science, Soochow University, Suzhou 215006, P. R. China}
\address[4]{School of Physics, Northwest University, Xi’an 710127, P. R. China}
\begin{abstract}
Patterns by self-organization in nature have garnered significant interest in a range of disciplines due to their intriguing structures. 
In the context of the snowdrift game (SDG), which is considered as an anti-coordination game, but the anti-coordination patterns are counterintuitively rare. 
In the work, we introduce a model called the Two-Agents, Two-Action Reinforcement Learning Evolutionary Game ($2\times 2$ RLEG), and apply it to the SDG on regular lattices. We uncover intriguing phenomena in the form of Anti-Coordinated domains (AC-domains), where different frustration regions are observed and continuous phase transitions at the boundaries are identified. 
To understand the underlying mechanism, we develop a perturbation theory to analyze the stability of different AC-domains. Our theory accurately partitions the parameter space into non-anti-coordinated, anti-coordinated, and mixed areas, and captures their dependence on the learning parameters.
Lastly, abnormal scenarios with a large learning rate and a large discount factor that deviate from the theory are investigated by examining the growth and nucleation of AC-domains. 
Our work provides insights into the emergence of spatial patterns in nature, and contributes to the development of theory for analysing their structural complexities.

\end{abstract}


\begin{highlights}
\item Anti-coordinated patterns naturally emerge in snowdrift game by reinforcement learning.
\item A perturbation theory is successfully developed 
\item The effect of memory and frustration greatly influence these patterns.
\end{highlights}



\end{frontmatter}


\section{Introduction}\label{introduction} 


The spatial patterns by self-organization are ubiquitous in ecosystems~\cite{rietkerk2008regular,meron2012pattern,dobramysl2018stochastic,reichenbach2007mobility,widder2016challenges}, physics~\cite{cross1993pattern,gollub1999pattern,lebon2008understanding,staliunas2002faraday,battiston2020networks}, biology~\cite{koch1994biological,wong2016dynamic,meinhardt1982models,Chen2017Fundamental} and other fields~\cite{walgraef2012spatio,resnik1997mathematics}. The emergence of large-scale intriguing patterns in space is rooted in the operation of positive feedback mechanisms at small scales that disrupt uniform states, which has been extensively studied in the field of nonlinear dynamics. As a pioneer, Turing first proposed the activator-inhibitor principle that relies on local feedback as an explanation for the pattern formation~\cite{turing1990chemical}. This principle has since been widely accepted to explain patterns observed across various disciplines~\cite{rovinsky1993self,klausmeier1999regular,sun2012pattern,Chen2019Persistent}. Along this direction, significant progress has been made in revealing various mechanisms underlying spatial self-organization: localized disturbance-recovery processes that give rise to power laws in cluster geometry~\cite{pascual2005criticality}, oscillating consumer-resource interactions leading to spiral waves~\cite{hassell1991spatial,thompson2012food}, as well as scale-dependent feedback resulting in regular patterns~\cite{levin1998ecosystems,preiser2018social}.
Another direction stems from non-equilibrium phase transitions~\cite{cross1993pattern,gollub1999pattern,lebon2008understanding,brown2017coarsening}, where several theoretical methods are also developed to analyze pattern formation~\cite{langer1980instabilities,sattinger1978group,cross1993pattern,gollub1999pattern,brown2017coarsening}. 

In parallel, over the past fifty years, evolutionary dynamics have been introduced and adopted as one of the main theoretical frameworks to investigate ecosystems and human behaviors~\cite{10.1093/biosci/biaa079,doi:10.1177/000169939203500302,rogalski2017human}. It is therefore natural to utilize this framework to explore the emergence of various patterns in both systems, where two approaches are generally utilized: continuum partial differential equations~\cite{neuhauser2001mathematical,holmes1994partial,turner2001landscape,cantrell2004spatial} and discrete agent-based models~\cite{packard1985two,esmaeili2018perturbing,szolnoki2014cyclic}. Some researchers focus on the analysis of general mechanisms, the stability, and diversity of these patterns~\cite{dobramysl2018stochastic,reichenbach2007mobility,esmaeili2018perturbing}. Meanwhile, other researchers pay attention to the emergence of patterns in specific contexts, such as cooperation~\cite{nowak1992evolutionary,nowak1993spatial} and fairness~\cite{lamba2013evolution,andre2011evolution}, among other emergent phenomena. With the booming of network science, researchers now have been broadening their focus from classical spatiotemporal systems to different types of networked systems~\cite{asllani2014turing,allen2017evolutionary,nakao2010turing}, and even systems on hypergraphs~\cite{carletti2020dynamical,carletti2022pattern,wang2023spatio}.

In classical agent-based models, agents evolve according to the predetermined rules, such as imitation, replication~\cite{nowak1992evolutionary,allen2017evolutionary,ma2023emergence}, or majority rule~\cite{lima2022diffusive,crokidakis2012impact}, among others~\cite{zhang2016controlling,huang2012emergence,wang2006memory}. Inspired by psychology and neuroscience, reinforcement learning (RL) has been gradually replacing these fixed rules as it captured the agents' adaptivity by learning~\cite{song2022reinforcement,wang2022levy,bertsekas2019reinforcement,zhang2020understanding}. Nonetheless, works in this regards are still in its infancy to understand puzzling phenomena, such as cooperation, fairness and resource optimalization~\cite{zheng2023optimal}, aiming to provide mechanistic explanations for them. Several recent studies have shown that RL better explains conditional cooperation and its temporal patterns~\cite{ezaki2016reinforcement,izquierdo2007transient,izquierdo2008reinforcement,masuda2011numerical,horita2017reinforcement}.

However, there is a knowledge gap of how spatial patterns could be understood within this new framework, and what's the underlying mechanisms in such agent-based systems~\cite{fan2022incorporating}. 
In particular, the snowdrift game (SDG) is a well-known anti-coordination game, the emergence of anti-coordinated patterns are expected through self-organization. Though, it is surprisingly rare to observe such patterns within the classical models imitation. Within the RL framework, most of studies monitor the evolution of cooperation and concentrate on the impact of various factors on the cooperative prevalence~\cite{zhang2012novel,NI20094856,jia2013evolution,xu2022enhanced}. 
Therefore, we are particularly interested in the following questions: \emph{Can anti-coordination patterns emerge in the evolutionary RL game framework? If so, what factors determine the types of patterns and formation processes? And can we develop a theory of pattern formation within this framework?} Addressing these questions is of paramount significance because it helps us understand the emergence of patterns from the learning perspective, and could be a cornerstone for developing theory for more complex scenarios.


This paper is structured as follows: In Sec.~\ref{sec:model}, we introduce a model called Two-Agents Two-Actions Reinforcement Learning Evolutionary Game ($2\times 2$ RLEG), which combines the Q-learning algorithm with the framework of evolutionary game theory. 
In Sec.~\ref{sec:results}, we apply the RLEG to SDG on three regular lattices and show the emergence of Anti-Coordinated domains (AC-domains). 
Upon the preliminary analysis in Sec.~\ref{subsec:preliminary}, we propose a perturbation theory to analyze the stability of these domains in Sec.~\ref{subsec:stability}. 
Using this theory, we partition the parameter space based on the emergence of AC-domains, and uncover potentially continuous phase transitions at their boundaries. 
In Sec.~\ref{subsec:dyamics_ACDs}, we investigate the evolution of AC-domains and uncover abnormal scenarios due to agents’ high learning rate and expectation of future. 
Lastly, in Sec.~\ref{subsec:composition}, we explore the composition of agents, which determines the cooperation preference within the population. 
Our conclusion and discussion are presented in Sec.~\ref{sec:conclusions}.



\section{A general reinforcement learning model} \label{sec:model}

Here, we present a general Two-player, Two-action Reinforcement Learning Evolutionary Game ($2\times 2$ RLEG) model in a structured population $\mathcal{N}$, where each agent plays the game with their nearest-neighbors following the Q-learning algorithm~\cite{10.1007/BF00992698}. 
Let's denote the action set $\mathcal{A}=\set{1, 0}$, where $1$ and $0$ respectively represent cooperation (C) and defection (D).  The state set for a given player $i$ is the Cartesian product between set of neighbors action combinations and action set, denoted as $\mathcal{S}^i=\set{00,10,\cdots,|\Omega^{i}|0,01,11,\cdots,|\Omega^{i}|1}$, where $\Omega^i$ is its neighborhood. In Q--learning algorithm, each player has a Q-table in hand to guide their choice of action, which is a matrix on Cartesian product for 
states  -- actions  $\mathcal{S}\times\mathcal{A}\rightarrow \mathbb{R}$
\begin{eqnarray}\label{eq:q_table}
	{\bf Q}^{i}(\tau) =
	\left(                 
	\begin{array}{cc}   
		Q^{i}_{00,0}(\tau) & Q^{i}_{00,1}(\tau)\\  
		Q^{i}_{10,0}(\tau) & Q^{i}_{10,1}(\tau)\\ 
		\vdots        & \vdots \\
		Q^{i}_{|\Omega_{i}|0,0}(\tau) & Q^{i}_{|\Omega_{i}|0,1}(\tau)\\
		Q^{i}_{01,0}(\tau) & Q^{i}_{01,1}(\tau)\\
		Q^{i}_{11,0}(\tau) & Q^{i}_{11,1}(\tau)\\ 
		\vdots       & \vdots \\
		Q^{i}_{|\Omega_{i}|1,0}(\tau) & Q^{i}_{|\Omega_{i}|1,1}(\tau) 
	\end{array}
	\right). 	\nonumber
\end{eqnarray}
The Q-value $Q^i_{s,a}$ for each item is the action-value function, measuring the value of the corresponding action $a$ for a given state $s$; a larger value of an item $Q^i_{s,a}>Q^i_{s,\hat{a}}$ means that the corresponding action $a$ is more preferred than the other $\hat{a}$. The Q-table can then be interpreted as the policy the player used to make their choice.

Initially, all players are randomly assigned C or D, and each items in their Q-tables are also randomly and independently initialised within the range of $[0,1]$.  
The evolution follows synchronous updating scheme. At each time step $\tau$, the state is computed according to $s^{i}(\tau) = n^{i}(\tau)a^{i}(\tau)$, where $n^{i}(\tau)=\sum_{j\in\Omega^i} a^{j}(\tau-1)$. The evolutionary protocol is divided into two processes: \emph{gaming} and \emph{learning} processes. 
In the gaming process, with probability $\epsilon$, an action is randomly chosen within $a^i(\tau)\in\mathcal{A}$. Otherwise, player $i$ takes action following the guidance of its Q-table
\begin{eqnarray}
	a^{i}(\tau) \rightarrow h({\bf Q}^i, s^i) = \arg \max\limits_{a'}\{Q^{i}_{s^i,a'}(\tau)\}, a'\in\mathcal{A},  \label{eq:action}
\end{eqnarray}
where $\arg\max\limits_{a'}\{Q^{i}_{s^i,a'}(\tau)\}$ is the action corresponding to the maximum Q-value in the row of state $s^i$. 
The parameter $0 < \epsilon \ll 1$ is the exploration rate, to introduce some randomness besides the exploitation of the Q-table.  
When all agents make their decisions, the rewards can then be computed according to the payoff matrix predefined by the game as
\begin{eqnarray}\label{eq:payoff_matrix}
	{\bf \Pi} =
	\left(                 
	\begin{array}{cc}   
		\Pi_{11} & \Pi_{10}\\  
		\Pi_{01} & \Pi_{00}\\  
	\end{array}
	\right) 
	= 	\left(                 
	\begin{array}{cc}   
		R & S\\  
		T & P\\  
	\end{array}
	\right) 
	. \nonumber
\end{eqnarray}
Here, $\Pi_{aa^{\prime}}$ denotes $i$'s payoff if $i$ with action $a^{i}(\tau) = a$ is against one of neighbours with the 
action $a^{j}(\tau)= a^{\prime}$. Specifically, the reward received by $i$ is 
\begin{eqnarray}\label{eq:reward}
	r^{i}(\tau) &=& \mathbbm{1}_{a_{i}=1}\left\{\sum\limits_{j\in\Omega_{i}}a^{j}\cdot\Pi_{11} + \sum\limits_{j\in\Omega_{i}}[1-a^{j}]\cdot\Pi_{10}\right\} + \nonumber\\
	& &\mathbbm{1}_{a_{i}=0}\left\{\sum\limits_{j\in\Omega_{i}}a^{j}\cdot\Pi_{01} + \sum\limits_{j\in\Omega_{i}}[1-a^{j}]\cdot\Pi_{00}\right\},
\end{eqnarray}
in which $\mathbbm{1}_{predicate}$ denotes the random variable that is $1$ if \emph{predicate} is true and 0 if it is not. 
 By then, the gaming process is then completed, and the new state for each player can be computed as $s'\equiv s^{i}(\tau+1) =(n^{i}(\tau+1), a^{i}(\tau))$, 
 where $n^{i}(\tau+1)=\sum_{j\in\Omega^i} a^{j}(\tau)$.
 

In the learning process, the element $Q^{i}_{s,a}$ is to revised as follows
\begin{eqnarray}\label{eq:update_Q}
	Q^{i}_{s,a}(\tau + 1) &= & g({\bf Q}^{i}(\tau),
	\bar{r}^{i}(\tau)) \nonumber\\
	&=&(1-\alpha)Q^{i}_{s,a}(\tau) +\alpha\left[\gamma Q_{s',a'}^{i\max}(\tau)+\bar{r}^{i}(\tau)\right].
\end{eqnarray}
Here, we adopt the average reward defined as $\bar{r}^{i}(\tau) = r^{i}(\tau)/|\Omega^{i}|$. $\alpha \in (0, 1]$ is the learning rate and $\gamma \in [0,1)$ is the discount factor. The former 
reflects the strength of memory effect since a large value of $\alpha$ means that the player is forgetful. 
The latter determines the importance of future rewards since $Q_{s',a'}^{\max}$ is the maximum element that one can expect within the new state $s'$. 
 
\begin{algorithm}[htbp!]  
	\caption{The Algorithm of Reinforcement Learning Evolutionary Game}\label{algorithm:protocol}
	\LinesNumbered 
	\KwIn{Learning parameters: $\alpha$, $\gamma$, $\epsilon$; Payoff matrix: $\Pi$; Population:$\mathcal{N}$}
	Initialization\;
	\For{$i$ in $\mathcal{N}$}{
		Pick an action randomly from $\mathcal{A}$\;
		Create a Q-table with each item in the matrix near zero\;
	}
	\For{$i$ in $\mathcal{N}$}{
		Generate state:$s^{i} \rightarrow n^{i}a^{i}$\;
	}
	\Repeat{\text the system becomes statistically stable or evolves for the desired time duration}
	{\For{$i$ in $\mathcal{N}$}{
			Generate a random number $p$\;
			\eIf{$p<\epsilon$}{
				Pick an action randomly from $\mathcal{A}$
			}{
				Take an action according to state, Q-table and Eq.~(\ref{eq:action})
			}
		}
		\For{$i$ in $\mathcal{N}$}{
			Get reward $\bar{r}_i$ according to their actions and payoff matrix $\Pi$\;
			Update Q-table according to Eq.~(\ref{eq:update_Q}) and state $s^{i} \rightarrow n^{i}a^{i}$\;
		}
	}
\end{algorithm}

The two processes repeat until the system becomes statistically stable or evolves for the desired time duration.
To summarize, the pseudo code is provided in Algorithm~\ref{algorithm:protocol}.

\section{Simulation results for snowdrift game}\label{sec:results}

In this work, players are placed on a regular lattice, where $|\Omega^{i}| = |\Omega|$ for any agent. They play SnowDrift Game (SDG)~\cite{sugden2004economics,smith1982evolution} with their nearest-neighbors within our RLRG framework. The payoff matrix for the SDG is as follows
\begin{eqnarray} \label{eq:payoff_matrix}
	{\bf \Pi} 
	=
	\left(
	\begin{array}{cc}
		R & S\\
		T & P \\
	\end{array}
	\right)
	=
	\left(
	\begin{array}{cc}
		1& 1-b \\
		1+b & 0 \\
	\end{array}
	\right),
\end{eqnarray}
where $b\in (0, 1)$ is a tunable game parameter.
Here we focus on the average cooperation preference $\langle\bar{f}_{c}\rangle$ of the population within the stable stage, defined as
\begin{eqnarray}\label{eq:cooperation_preference}
	\langle\bar{f}_{c}\rangle := \sum_{\tau=t_{0}}^{t}\frac{\sum_{i\in\mathcal{N}}\mathbbm{1}_{a^{i}(\tau)=1}}{(t-t_{0})|\mathcal{N}|},
\end{eqnarray}
in which $t$ is the overall steps used for simulations and $t_{0}$ denotes the transient time step.
In this study, we investigate the cooperation preference in different types of lattice, and examine the impact of the game parameter $b$ and the learning parameters $\alpha$ and $\gamma$. 


\begin{figure}[htbp!]
	\centering
	\includegraphics[width=0.75\textwidth]{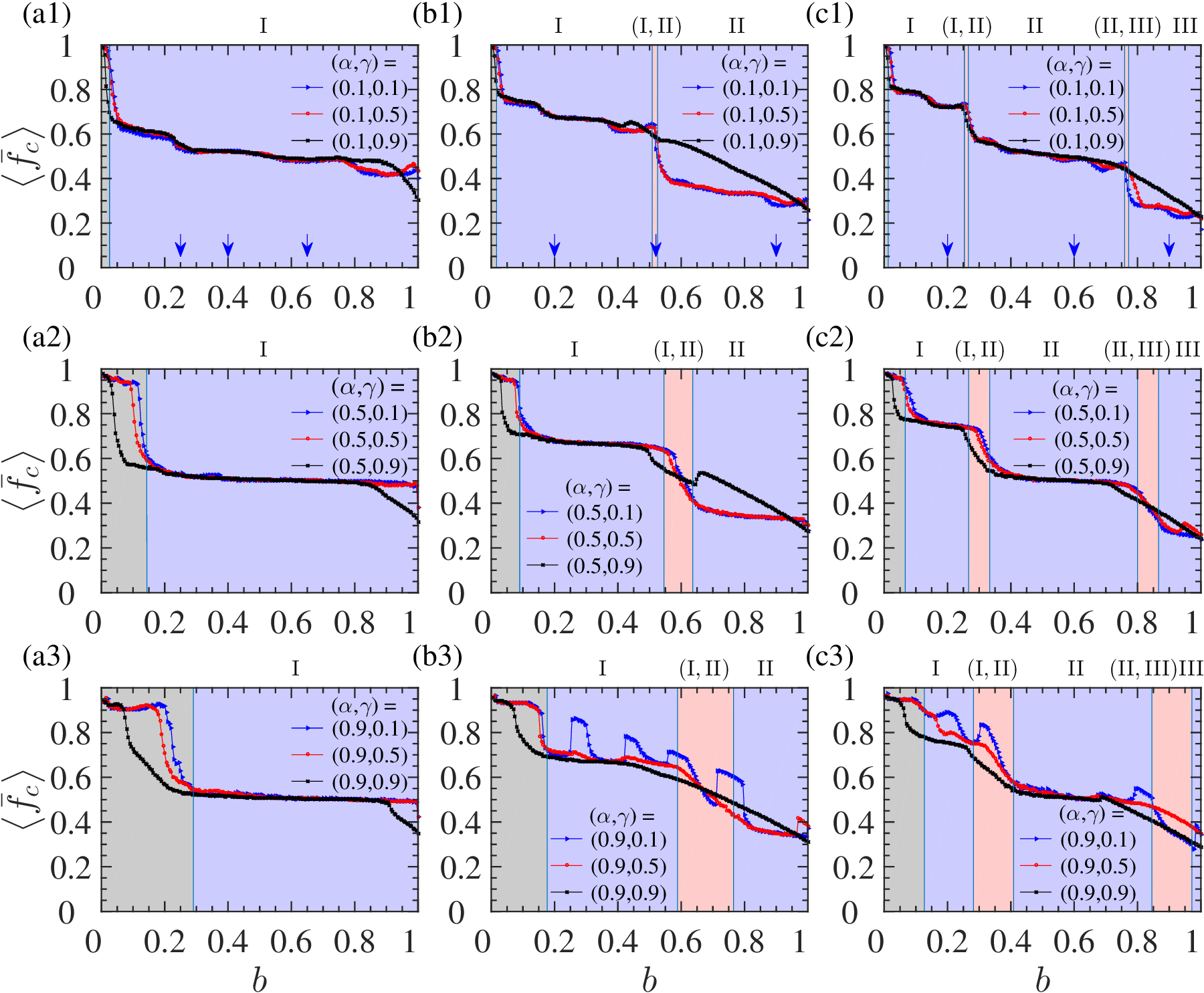}
	\caption{{\bf The average cooperation preference as the function of game parameter $b$ on three different regular lattices and under combinations of learning parameters.} In (a1-a3), (b1-b3) and (c1-c3), the regular lattices are von Neumann lattices ($|\Omega| =4$), triangular lattices ($|\Omega| =6$) and Moore lattices ($|\Omega| = 8$) respectively.
Theoretically predicted Anti-Coordinated areas (AC-areas), Non-Anti-Coordinated areas (NAC-areas) and Mixed areas (M-areas) are depicted in blue, gray, and orange shading, respectively. In the top panels, we identify and label the type of Anti-Coordinated domain (AC-domain) in the corresponding AC-area, as well as the types of coexisting AC-domains observed in the corresponding M-area.
 Other parameters: $\epsilon =0.01$, and the population size $|\mathcal{N}|=10000$.
	} 
	\label{fig:b} 
\end{figure}

We first fix different learning parameter combination$\tuple{\alpha, \gamma}$ and calculate $\langle\bar{f}_{c}\rangle$ as the function of the game parameter $b$ for three types of lattices, as shown in Fig.~\ref{fig:b}. These include the von Neumann lattice [Fig.~\ref{fig:b}(a1-a3)], the triangular lattice [Fig.~\ref{fig:b}(b1-b3)], and the Moore lattice [Fig.~\ref{fig:b}(c1-c3)]~\cite{flores2022cooperation,PhysRevE.72.047107}, with the periodic boundary condition being used in all cases. 
Fig.~\ref{fig:b} shows a decline of $\langle\bar{f}_{c}\rangle$ as $b$ is increased, which is in line with our expectation. But interestingly, there are also flat regions, where the decrease of cooperation prevalence becomes very slow as $b$ is varied.
Detailed examination shows that the number, position, and volatility of these flat regions are influenced by both the type of lattice and the learning parameter combination of $\tuple{\alpha, \gamma}$. These include: (1) a larger degree $|\Omega|$ yields more flat regions; (2) the learning rate $\alpha$ is found to have more significant impact on the position of flat regions than $\gamma$ for the given lattice type; (3) a larger $\alpha$ generally narrows the flat regions and makes them more volatile.

\begin{figure}[htbp!]
	\centering
	\includegraphics[width=0.75\textwidth]{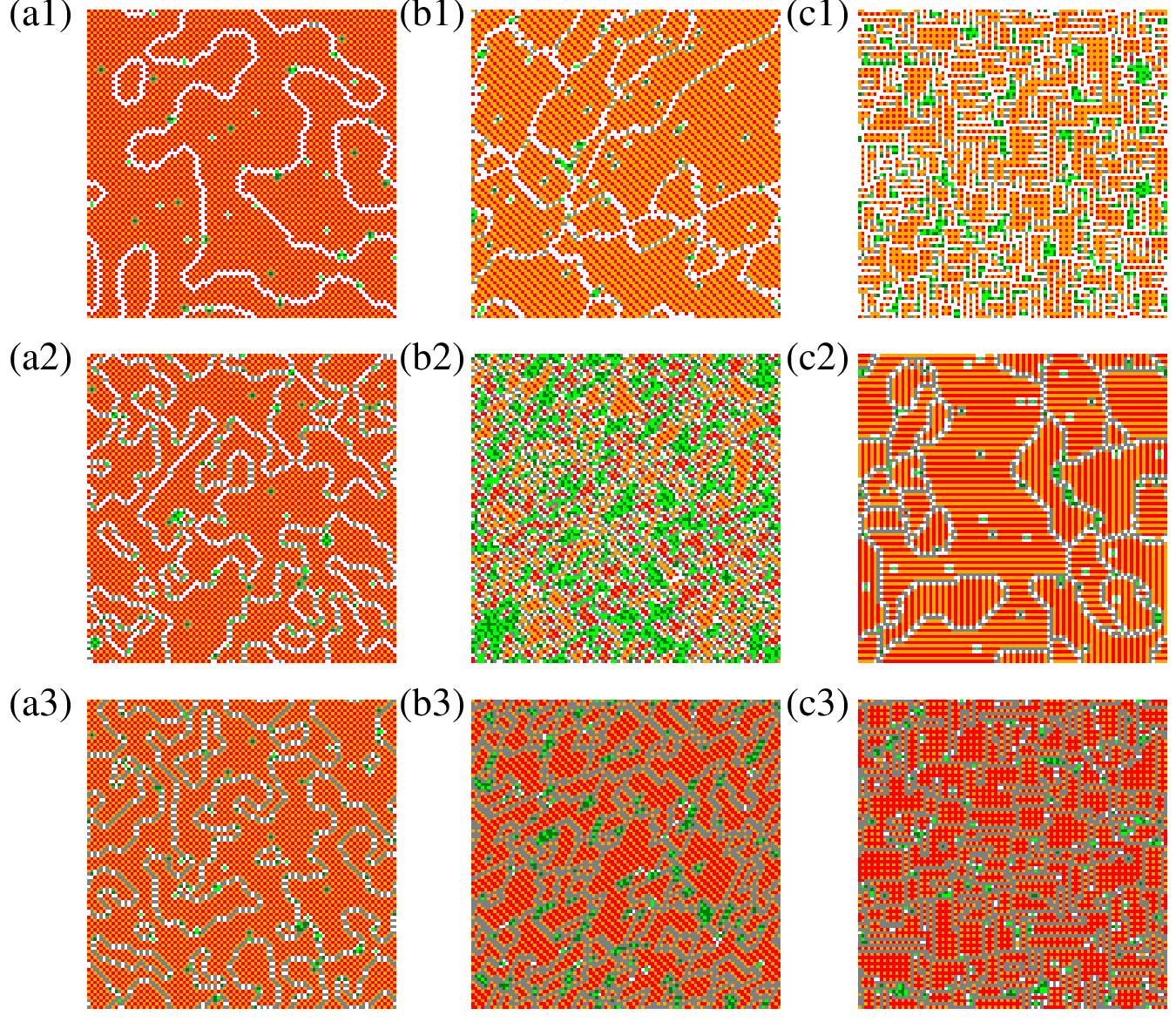}
	\caption{{\bf Typical snapshots on three lattices when the evolution becomes stable.} 
	The snapshots of actions in (a1-a3) are on the von Neumann lattices, with game parameter being $b = 0.25, 0.4, 0.65$ respectively. 
	The snapshots of actions in (b1-b3) are on the triangular lattices, with game parameter being $b = 0.2, 0.52, 0.9$respectively. 
	The snapshots of actions in (c1-c3) are on the Moore lattices, with game parameter being $b = 0.2, 0.65, 0.9$ respectively.
        The entire domains for a snapshot is made up Anti-Coordinated domains (AC-domains) enclosed by domain walls, and disordered domains (D-domains). 	
        Here the cooperators and defectors are encoded by \domainc{} and \domaind{} within the AC-domains, by \wallc{} and \walld{} on the walls, and by \disorderc{} and \disorderd{} within the D-domains,  respectively.  
        Parameters: $\tuple{\alpha,\gamma} = \tuple{0.1, 0.1}$, $\epsilon = 0.01$, the population size $|\mathcal{N}| = 10000$.} 
	\label{fig:snapshots} 
\end{figure}

To develop some intuition, we provide some typical snapshots as shown in Fig.~\ref{fig:snapshots}. 
In most panels, we observe that locally stable domains are enclosed by ``{\em domain walls}''; Within the locally stable domains, referred to as ``{\em anti-coordinated domains}'' (AC-domains), the action preference is opposite to that of their neighbors, in analogous to spin configurations in the anti-ferromagnetic system. 
 Due to the different topologies of different lattices, the unit cells $\mathcal{U}$ that make up AC-domains are different. But even on the same topology, the configurations in AC-domains are still could be different [see Fig.~\ref{fig:snapshots}(c1-c3)], and we refer to as {\em ``Anti-Coordinated areas''} (AC-areas) for distinction.
 These snapshots immediately explain why flat regions are seen in Fig.~\ref{fig:b}. This is because the variation in parameter $b$ only shift the domain walls whereas the configurations of AC-domains remain largely unchanged.
 

Different from the von Neumann lattice, there are triangular motifs in the triangular and Moore lattices, which interestingly result in frustration in AC-domains due to anti-coordination [see Fig.~\ref{fig:snapshots} (b1-b3) and (c1-c3)]. 
And, the frustrated agents in AC-domains prefer cooperation for low $b$, but this choice is gradually replaced by defection as $b$ increases. The flip of the preference brings the diversity of AC-domains and the emergence of AC-areas on triangular and Moore lattices as shown. 


Besides AC-domains, we also observe localized \emph{``Disordered domains''} (D-domains) [see Fig.~\ref{fig:snapshots} (b2)], which is in between two AC-areas of Fig.~\ref{fig:b}.
For the combination of parameters between two AC-areas of Fig.~\ref{fig:b}, we observe the presence of coexisting AC-domains, along with localized D-domains [see Fig.~\ref{fig:snapshots} (b2)].  And, the coexisting AC-domains are the dominating structure in the two AC-areas, respectively. We call these areas as \emph{``Mixed areas''} (M-areas) because domains within their snapshot consistently exhibit mixed. While the area, where cooperation preference in the population is high, apart from the AC-areas and M-areas, is appoint to the \emph{``Non-Anti-Coordinated areas''} (NAC-areas). 
In sum, there are three types of configurations in the domain, namely NAC-area, AC-areas, and M-areas. And there are subclass of AC-areas, depending on their configuration of AC-domains.

\section{Mechanism analysis}\label{sec:analysis}
\subsection{The preliminary}\label{subsec:preliminary}
\begin{figure}[htbp!]
	\centering
	\includegraphics[width=0.75\textwidth]{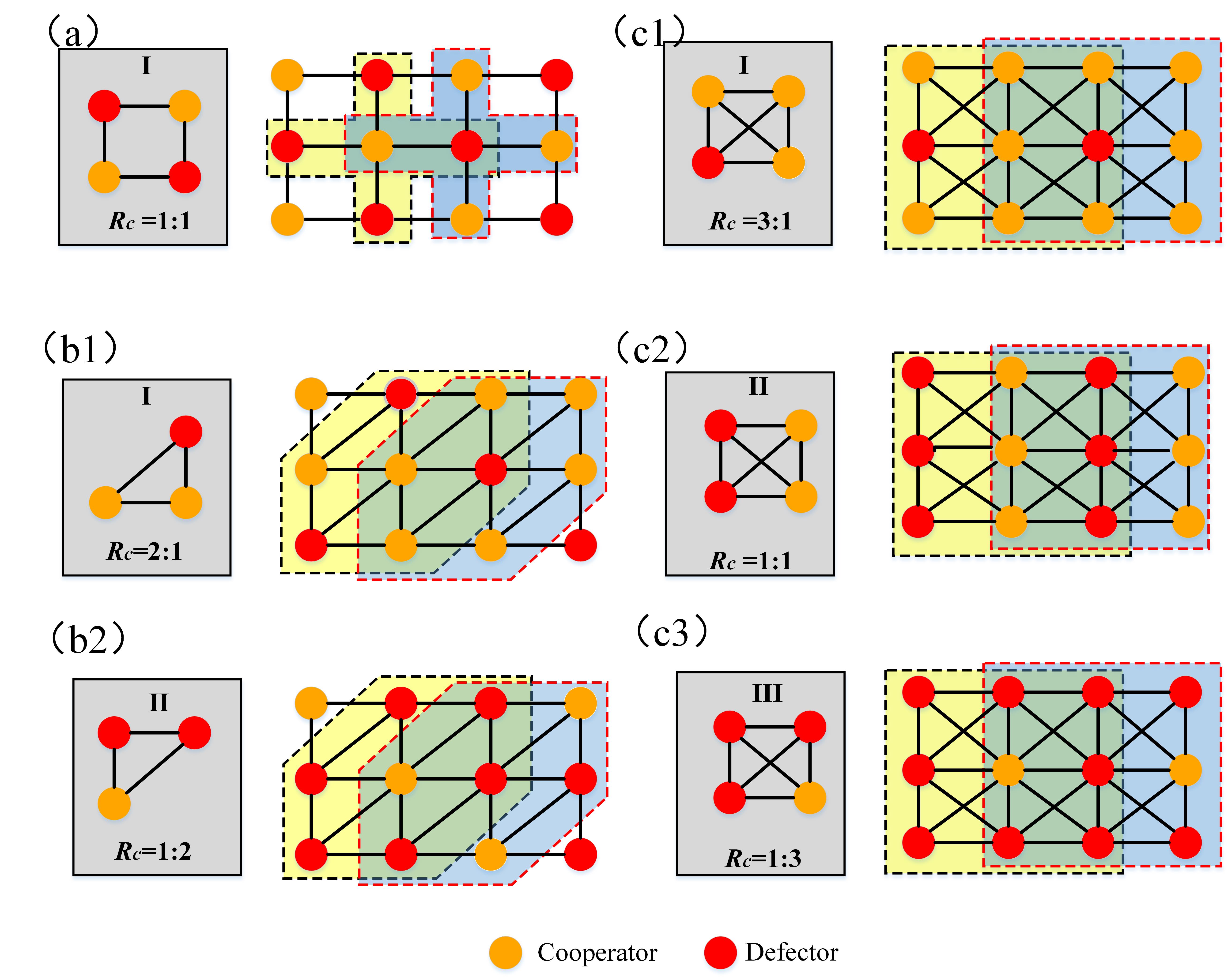}
	\caption{
	{\bf The unit cells and basic blocks of anti-coordinated domains on different lattices.} 
	(a) shows the unique basic unit cell on the von Neumann lattice, with the ratio cooperators to defector $R_{c} =1:1$, appointed to type-\RNum{1}. 
	(b1-b2) give the two basic unit cells on the triangular lattice, with $R_{c} = 2:1$ and $1:2$, appointed to type-\RNum{1} and -\RNum{2}.  
	(c1-c3) display three basic unit cells on the Moore lattice, with $R_{c} = 3:1$, $1:1$, and $1:3$, appointed to type-\RNum{1}, -\RNum{2}, and -\RNum{3}. 
	Here, the cooperator-centred block $\mathcal{B}_{c}$ and defector-centred block $\mathcal{B}_{d}$ are indicated with blue and yellow shadows, respectively.
	} 
	\label{fig:cell} 
\end{figure}

In the subsection, we present a brief preliminary as the foundation for the following subsections. The flat regions and different types of AC-domains in Fig.~\ref{fig:snapshots} are due to the presence of unit cell, see Fig.~\ref{fig:cell}. In Fig.~\ref{fig:cell}(a), (b1-b2), and (c1-c3), we present the unit cell for each domain type on the von Neumann lattice, triangular lattice, and Moore lattice, respectively. For a given type of unit cell, one can define the ratio of the number of cooperators to the number of defectors within it,
\begin{eqnarray}
	R_{c}(\mathcal{U}_{X}) :=\frac{\sum_{i\in\mathcal{U}_X}\mathbbm{1}_{a^{i}=1}}{\sum_{i\in\mathcal{U}_X}\mathbbm{1}_{a^{i}=0}},  \nonumber
\end{eqnarray}
which reflects the cooperation preference in this unit. 


Obviously, the action of a given individual is influenced by the actions of their neighbors in the cell, which also influences the stability of the engaged AC-domains. 
Here we designate $\mathcal{B}_{c}$ and $\mathcal{B}_{d}$ for the central agents in the blocks are $1$ (cooperator) and $0$ (defector), respectively.
Together with its neighbors, they form the basic blocks.
In the following study, we are going to investigate how the perturbations arising from agents’ exploration affect the stability of basic blocks.


\subsection{The partitioning of parameter space}\label{subsec:stability}

The above simulations illustrate that the parameter space can be partitioned into NAC-area, AC-areas, and M-areas, based on the character of the snapshots. And, the locations of AC-areas play a crucial role in determining the locations of other areas. Hence, in the subsequent analysis, we prioritize our attention towards examining the location of different AC-areas in space. 

For any AC-area, there is always a specific type of stable AC-domain. The stability of this AC-domain is evidently dependent on the stability of its constituent internal basic blocks $\mathcal{B}_{c}$ and $\mathcal{B}_{d}$. 
Hence, our principal focus is to investigate the necessary conditions that $\mathcal{B}_{c}$ and $\mathcal{B}_{d}$  must satisfy to attain internal stability during exploration. These conditions also dictate the location of the AC-area in parameter space. In our analysis, we appoint an agent’s exploratory action as the action that differs from the action obtained from its Q-table, i.e., $a(\tau) = 1 - h({\bf Q}, s)$. The stability conditions for $\mathcal{B}_{c}$ and $\mathcal{B}_{d}$ require that the structure of any agent’s Q-table remains intact and is not disrupted by the perturbation from its neighboring agents exploratory actions. In other words, the stability of these blocks is contingent upon the preservation of Q-table structure under the influence of neighboring exploratory actions.

Without loss of generality, we investigate two adjacent blocks $\mathcal{B}_{c}$ and $\mathcal{B}_{d}$ within any specific type-$X$ AC-domain. In this context, we assign the central cooperator in $\mathcal{B}_{c}$ to ``$\tilde{1}$'' and the central defector in $\mathcal{B}_{d}$ to ``$\tilde{0}$''. And for $\tilde{1}$ and $\tilde{0}$, the number of cooperators in their neighbours are $m$ and $n$, respectively. In our analysis, we make the assumption that
$\mathcal{B}_{c}$ and $\mathcal{B}_{d}$ have achieved stability at $\tau$. By making this assumption, we are able to establish a self-consistent relation for stability that in turn enables us to derive condition on stability. 

\begin{figure}[htbp!]
	\centering
	\includegraphics[width=0.75\textwidth]{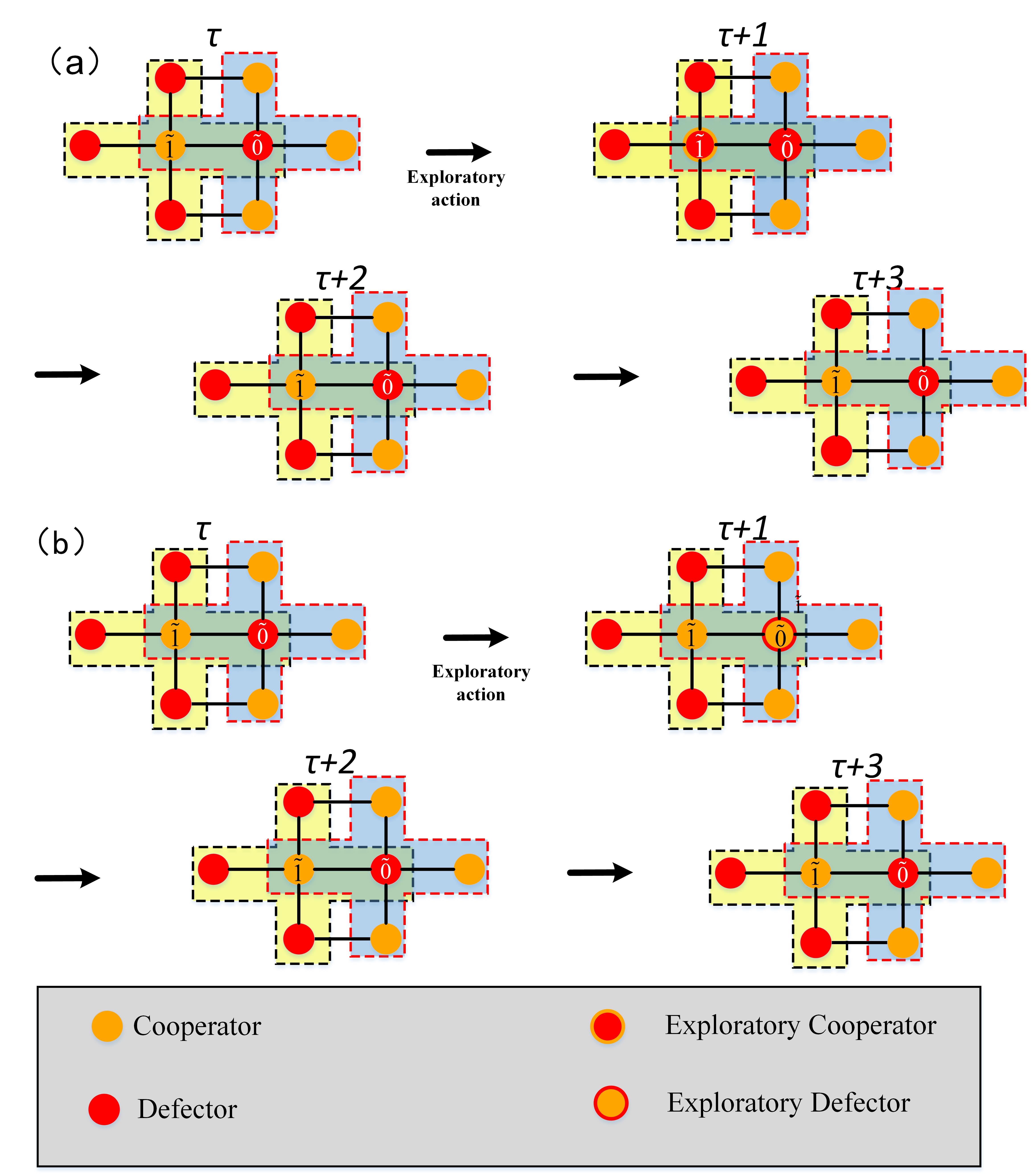}
	\caption{{\bf Schematic for the assumption to $1$-step recovery of basic blocks after exploratory actions.} In (a) and (b), both blocks $\mathcal{B}_{c}$ (highlighted with a yellow shadow) and $\mathcal{B}_{d}$ (highlighted with a blue shadow) has become stable at $\tau$ step. At $\tau+1$, the central agent ``$\tilde{1}$'' or ``$\tilde{0}$'' takes an exploratory action, but neither configuration of $\mathcal{B}_{c}$ and $\mathcal{B}_{d}$ can be broken by this action in the future. As a result, both configurations will be recovered at $\tau+2$ and sustained at $\tau+3$ in the assumption of the 1-step recovery. In the schematic, the exploratory action for $\tilde{1}$
		is represented by a red circle with an orange border denoting defection, while the exploratory action for $\tilde{0}$ is represented by an orange circle with a red border denoting cooperation.
	} 
	\label{fig:stability} 
\end{figure}

Given that $\mathcal{B}_{c}$ and $\mathcal{B}_{d}$ are stable, we get the stable Q-elements for $\tilde{1}$ and $\tilde{0}$ in $\mathcal{B}_{c}$ and $\mathcal{B}_{d}$ at $\tau$ that are 
\begin{eqnarray}
	\left\{
	\begin{array}{l}
		Q^{\tilde{1}}_{m1,1}(\tau) = Q^{\tilde{1}*}_{m1,1} = \displaystyle{\frac{m\Pi_{11}+\left(|\Omega|-m\right)\Pi_{10}}{|\Omega|\left(1-\gamma\right)}} \\
		Q^{\tilde{0}}_{n0,0}(\tau) = Q^{\tilde{0}*}_{n0,0} = \displaystyle{\frac{n\Pi_{01}+\left(|\Omega|-n\right)\Pi_{00}}{|\Omega|\left(1-\gamma\right)}} 
	\end{array}
	\right.
\end{eqnarray}
Additionally, in terms of stability, it is imperative that any exploratory actions taken by either
$\tilde{1}$ or $\tilde{0}$ at $\tau + 1$
do not disrupt the structure of their own Q-tables, as well as the Q-tables of the central agent $\tilde{0}$'s or $\tilde{1}$'s of their adjacent block in the future. Rather, the original configurations of $\mathcal{B}_{c}$ and $\mathcal{B}_{d}$ are expected to be recovered over steps, even if initially disrupted by the exploratory action of $\tilde{1}$ or $\tilde{0}$. Here, we additionally propose an assumption, $1$-step recovery, stating that following their respective exploratory actions, the original configurations of $\mathcal{B}_{c}$ and $\mathcal{B}_{d}$ will be recovered at the next step $\tau + 2$ [see Fig.~\ref{fig:stability}]. Then, the Q-element associated with $\tilde{1}$'s or $\tilde{0}$'s exploratory action at $\tau+1$ and $\tau+2$ can be obtained as follows 
\begin{eqnarray}
	\left\{
	\begin{array}{l}
		Q^{\tilde{1}*}_{m1,0} = \displaystyle{\frac{m\Pi_{01}+\left(|\Omega|-m\right)\Pi_{00}}{|\Omega|}}+\gamma Q^{\tilde{1}*}_{m0,1} \\ 
		Q^{\tilde{0}*}_{n0,1} = \displaystyle{\frac{n\Pi_{11}+\left(|\Omega|-n\right)\Pi_{10}}{|\Omega|}}+\gamma Q^{\tilde{0}*}_{n1,0} \\
	\end{array}
	\right.
\end{eqnarray}
and 
\begin{eqnarray}
	\left\{
	\begin{array}{l}
		Q^{\tilde{1}*}_{m0,1} = \displaystyle{\frac{m\Pi_{11}+\left(|\Omega|-m\right)\Pi_{10}}{|\Omega|}}+\gamma Q^{\tilde{1}*}_{m1,1} =  Q^{\tilde{1}*}_{m1,1}\\
		Q^{\tilde{0}*}_{n1,0} = \displaystyle{\frac{n\Pi_{01}+\left(|\Omega|-n\right)\Pi_{00}}{|\Omega|}}+\gamma Q^{\tilde{0}*}_{n0,0} =  Q^{\tilde{0}*}_{n0,0}
	\end{array}
	\right.
\end{eqnarray}

Furthermore, the assumption to $1$-step recovery lead us to conclude that the exploratory action executed by either $\tilde{1}$ or $\tilde{0}$ 
will also not break the structure of the Q-tables belonging to their neighboring agents, $\tilde{0}$ and $\tilde{1}$ at $\tau+1$, respectively. 
Then, we have 
\begin{eqnarray}
	\left\{
	\begin{array}{l}
		\begin{aligned}
			&Q^{\tilde{0}}_{n0,0}(\tau+1)  =  (1-\alpha)Q^{\tilde{0}*}_{n0,0} + \alpha\left[\gamma Q^{\tilde{0}*}_{(n-1)0,0} + \right.  \\
			&\quad \left. \displaystyle{\frac{(n-1)\Pi_{01} +(|\Omega|-n+1)\Pi_{00}}{|\Omega|}}\right] \geq Q^{\tilde{0}*}_{n0,1}     \\
			&Q^{\tilde{1}}_{m1,1}(\tau+1)  =  (1-\alpha)Q^{\tilde{1}*}_{m1,1} + \alpha\left[\gamma Q^{\tilde{1}*}_{(m+1)1,1} + \right.  \\
			&\quad \left. \displaystyle{\frac{(m+1)\Pi_{11} +(|\Omega|-m-1)\Pi_{10}}{|\Omega|}}\right] \geq Q^{\tilde{1}*}_{m1,0},
		\end{aligned}
	\end{array}\label{eq:constraints_1}
	\right.
\end{eqnarray}
which guarantee the reconstructed blocks $\mathcal{B}_d$ and $\mathcal{B}_c$ will be sustained in the future.
Here, the $Q^{\tilde{0}*}_{(n-1)0,0}$ and $Q^{\tilde{1}*}_{(m+1)1,1}$ can be gained through our assumptions that are 
\begin{eqnarray}
	\left\{
	\begin{array}{l}
		Q^{\tilde{1}*}_{(n+1)1,1} = \displaystyle{\frac{n\Pi_{11}+\left(|\Omega|-n\right)\Pi_{10}}{|\Omega|}}+\gamma Q^{\tilde{1}*}_{n1,1} =  Q^{\tilde{1}*}_{n1,1}\\
		Q^{\tilde{0}*}_{(m-1)0,0} = \displaystyle{\frac{m\Pi_{01}+\left(|\Omega|-m\right)\Pi_{00}}{|\Omega|}}+\gamma Q^{\tilde{1}*}_{m0,0} =  Q^{\tilde{0}*}_{m0,0}  \\
	\end{array}\label{eq:constraints_2}
	\right.
\end{eqnarray}
In the Fig.~\ref{fig:stability}, we present a schematic representation illustrating the schematic for the assumption to $1$-step recovery for $\mathcal{B}_c$ and $\mathcal{B}_d$ on von Neumann networks according to the updating protocol. 

For the SDGs, the payoff matrix meets $\Pi_{01}>\Pi_{11}>\Pi_{10}>\Pi_{00}$. By substituting it into Eqs.~(\ref{eq:constraints_1}-\ref{eq:constraints_2}), we obtain the conditions to the stability for $\mathcal{B}_c$ and $\mathcal{B}_d$ that are
\begin{eqnarray}
	\left\{
	\begin{array}{l}
		\alpha \geq \displaystyle{\frac{m(\Pi_{01}-\Pi_{11}+\Pi_{10}-\Pi_{00})+|\Omega|(\Pi_{00}-\Pi_{10})}{\Pi_{11}-\Pi_{10}}} \\
		\alpha \leq \displaystyle{\frac{n(\Pi_{01}-\Pi_{11}+\Pi_{10}-\Pi_{00})+|\Omega|(\Pi_{00}-\Pi_{10})}{\Pi_{01}-\Pi_{00}}} 
	\end{array}
	\right.
\end{eqnarray}
Our findings suggest that, given our assumption, the conditions are contingent upon both $\alpha$ and $\bf{\Pi}$, but not upon $\gamma$. The results are similar to previous studies, which suggest that the specific type of AC-domain that emerges depends on the memory of the agents~\cite{wang2006memory}. 
Under our game settings, the conditions for a given $\alpha$ can be further rewritten as   
\begin{equation}
	\frac{|\Omega|-n+\alpha}{|\Omega|-\alpha}=b^{*}_{l} \leq  b \leq b^{*}_{u} = \frac{|\Omega|-m}{|\Omega|-\alpha} 
	\label{eq:constraints_final}
\end{equation}
with $b\in (0,1)$, $\alpha\in (0,1)$ and $m\le n$. Here, the lower limit $b^{*}_{l}$ and the upper limit $b^{*}_{u}$ correspond to the minimum and maximum game parameter required to maintain stability in the type-$X$ AC-domain, given a specified $\alpha$. The area between $b^{*}_{l}$ and $b^{*}_{u}$ in Eq.~(\ref{eq:constraints_final}) is also the type-$X$ AC-areas in parameter space.

For an M-area between type-$X$ and type-$X^{\prime}$ AC-areas, the parameters in space meet 
\begin{eqnarray}
	\frac{|\Omega|-m}{|\Omega|-\alpha}< b < \frac{|\Omega|-n^{\prime}+\alpha}{|\Omega|-\alpha}
	\quad\text{with} \quad n^{\prime} = m \label{eq:constraints_MA}
\end{eqnarray}
in which $m$ denotes the number of neighboring cooperators surrounding the central defector in type-
$X$ AC-domain, while $n^{\prime}$ denotes the number of neighboring cooperators surrounding the central defector in type-$X^{\prime}$ AC-domain. Based on Eq.~(\ref{eq:constraints_MA}), we observe that in the M-area context with $b^{*}_{l}< b \nleq b^{*}_{u}$ and $b^{\prime*}_{l}\nleq b < b^{\prime*}_{u}$, type-$X$ AC-domain remains stable when the cooperators take exploratory actions but becomes unstable when the defectors take exploratory actions. Conversely, type-$X^{\prime}$ AC-domains remains stable when the defectors take exploratory actions but becomes unstable when the cooperators take exploratory actions.  
Therefore, we refer to both type-$X$ and type-$X^{\prime}$
AC-domains as \emph{``semi-stable'' } due to their dynamics under exploratory actions. However, it’s worth noting that under the exploratory action of cooperators, the instability of type-$X$ increases with increasing game parameter $b$. While, under the exploratory actions of defectors, the instability of type-$X^{\prime}$ decreases with increasing $b$. 

\begin{figure}[htbp!]
	\centering
	\includegraphics[width=0.75\textwidth]{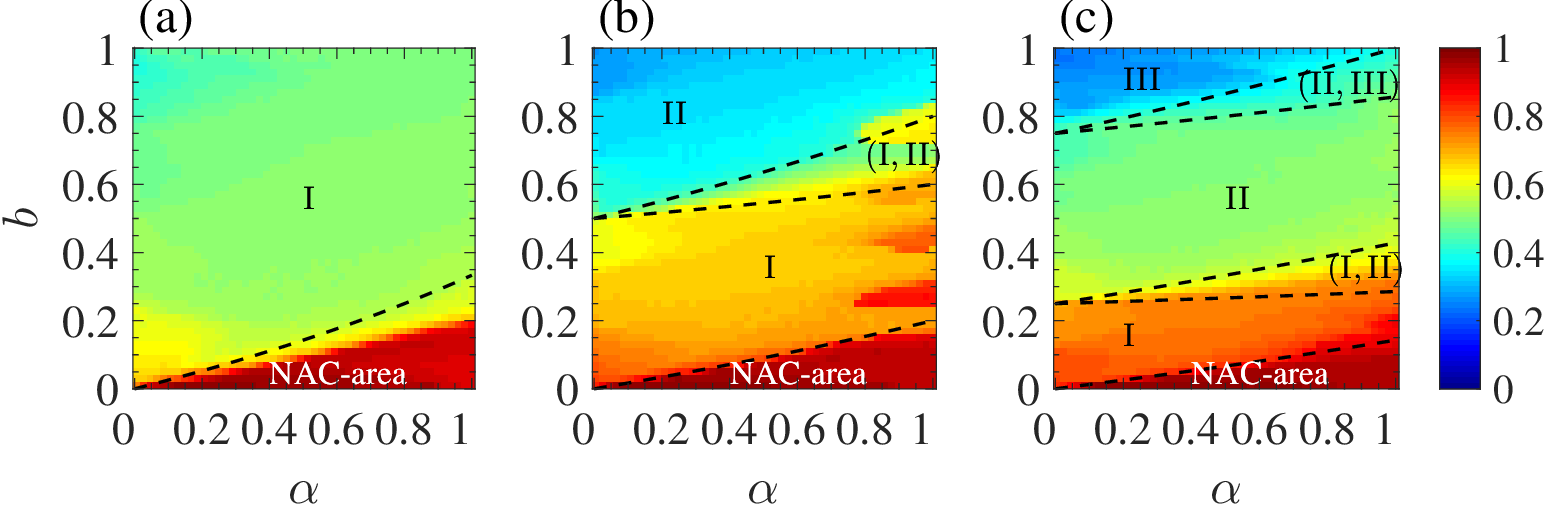}
	\caption{{\bf The cooperation preference by simulations and boundaries predicted by theory.} 
		The cooperation preference $\langle \bar{f}_{c}\rangle$ is color-coded in $\alpha-b$ space and the dashed lines are the theoretical boundaries of different \emph{Anti-Coordinated Areas} (AC-areas) and \emph{Non-Anti-Coordinated Areas} (NAC-area), (a-c) are for von Neumann lattice, triangular lattice and Moore lattice, respectively. 
		The types of AC-domains in AC-areas and the types of semi-stable AC-domains in M-areas are labeled in the corresponding areas. 
		These areas are delineated by dashed lines marking $b_{l}^{*}$ and $b_{r}^{*}$ in Eq.~(\ref{eq:constraints_final}). 
		Other parameters: $\gamma=0.5$, $\epsilon =0.01$ and $|\mathcal{N}| = 10000$.
	} 
	\label{fig:alpha_b} 
\end{figure}

Based on the analysis provided above, we can classify the domain wall and D-domains collectively as ``NonAC-domains'', denoted as $\hat{\mathcal{D}}$. Then, the cooperation preference in the population is 
\begin{eqnarray}
	\langle \bar{f}_{c}\rangle = \sum_{X} p(\mathcal{D}_{X}) f_{c}(\mathcal{D}_{X}) + p(\hat{\mathcal{D}}) f_{c}(\tilde{D})
	\label{eq:fc_analysis}
\end{eqnarray}
Here, $p(\mathcal{D}_{X})$ and $p(\hat{\mathcal{D}})$ are the proportions of type-$X$ AC-domains and NonAC-domains respectively, and meet $\sum_{X} p(\mathcal{D}_{X})+ p(\hat{\mathcal{D}}) = 1$. Additionally, $f_{c}(\mathcal{D}_{X}) = R_{c}(\mathcal{U}_{X})/(1+R_{c}(\mathcal{U}_{X}))$ and $f_{c}(\tilde{D})$ are the cooperation preferences for the agents belonging to the corresponding domains. So, it can be inferred that the cooperation preference $\langle \bar{f}_{c}\rangle$ for type-$X$ AC-area approximates $f_{c}(\mathcal{D}_{X})$. Hence, the cooperation preference $\langle \bar{f}_{c}\rangle$ can be utilized in the $\alpha-b$ space to evaluate the validation of our analysis as Figs.~\ref{fig:alpha_b} and ~\ref{fig:b} shows. The outcomes confirm that the theoretical partitioning hold true in a broad sense, encompassing von Neumann lattices, triangular lattices, and Moore lattices in $\alpha-b$.
Besides, the results in Fig.~\ref{fig:alpha_b} and Eqs.~(\ref{eq:constraints_final}-\ref{eq:constraints_MA}) indicate that high $\alpha$ leads to a narrowing of AC-areas but
an expansion of M-areas, similar to the observations in Figure~\ref{fig:b}.

\subsection{The evolution of anti-coordinated domain}\label{subsec:dyamics_ACDs}
In the previous analysis, we focus exclusively on the stability of the AC-domain, assuming that it is already formed. 
Yet, similar to the process of crystallization, the formation of an AC-domain consists of two distinct processes: nucleation and growth. There, it is also possible for the nucleation fail to develop, or fail to grow when in a competition with the D-domains. 
In our study, we will refer to the case as the ``\emph{normal scenario}'' when the composition of the domains is consistent with our theoretical predictions, and when the composition contradicts our theory, we term it ``\emph{abnormal scenario}''. 
In order to monitor the evolution of nucleation and growth, we present time series illustrating the number of AC-domains and their average sizes in both normal and abnormal scenarios in the following.

\subsubsection{Normal Scenarios}\label{subsec:normal_scenarios}
\begin{figure}[htbp!]
	\centering
	\includegraphics[width=0.85\textwidth]{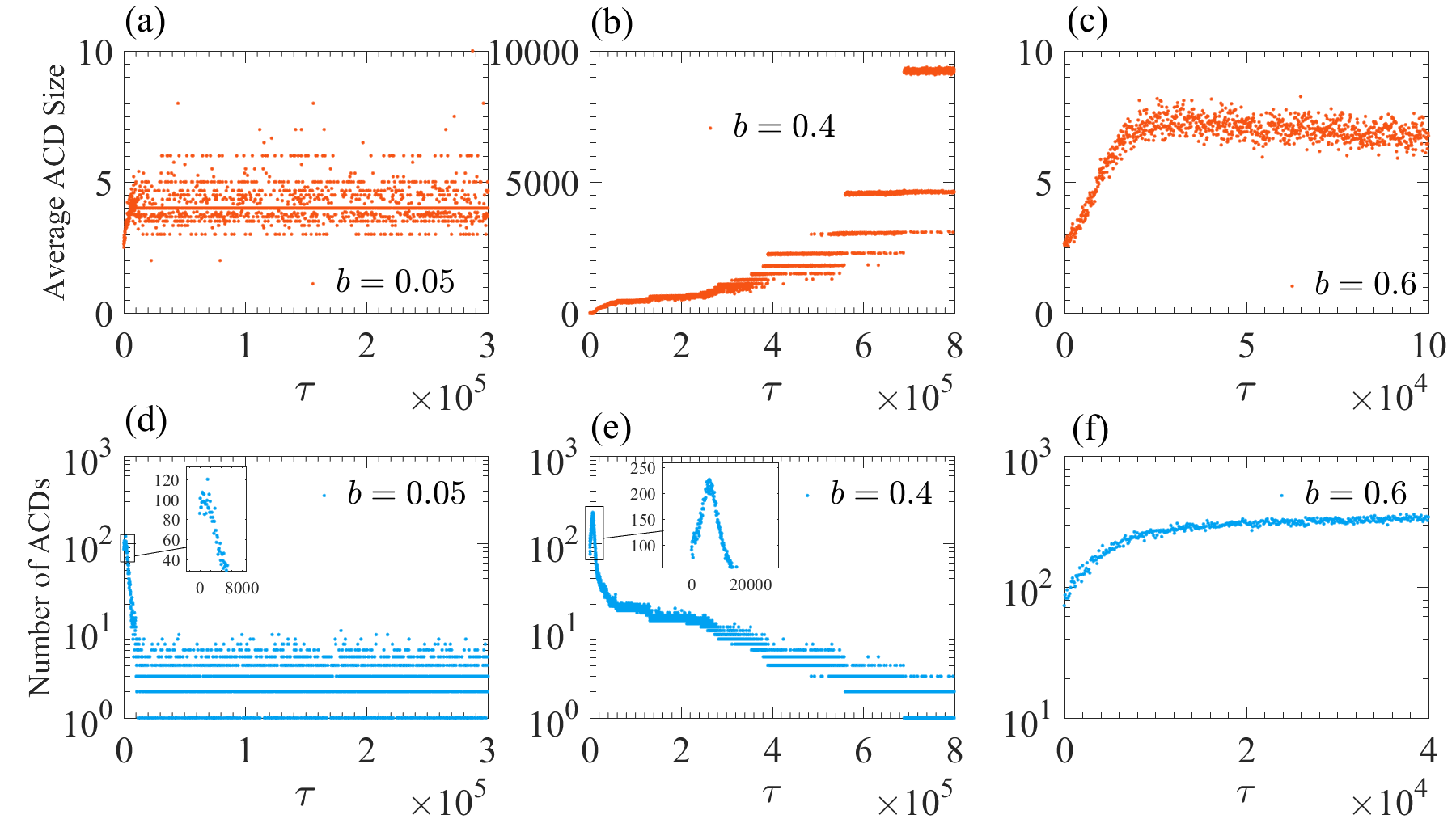}
	\caption{{\bf Statistics of anti-coordinated domains in the normal scenario on a triangular lattice.} We show the time series of the number and the average size of AC-domains within the NAC-area, the AC-area, and M-area, respectively shown in the left, middle and right panels.  
	Parameters: $\tuple{\alpha, \gamma} = (0.5, 0.1)$, $|\mathcal{N}| = 10000$.
	} 
	\label{fig:ts_normal} 
\end{figure}
In the NAC-area, the number of AC-domains decreases over time and eventually stabilizes at a very low level, as shown in the left panels of Fig.~\ref{fig:ts_normal}(d). Meanwhile, the average size of the AC-domains shows a slight increase but remains largely unchanged [Fig.~\ref{fig:ts_normal}(a)]. This means that the AC-domain nucleus fail to grow against invasion and become extremely scarce and nearly invisible in the system.
By contrast, in the AC-areas, the number of AC-domains increases initially and then decreases, while the average size of AC-domains steadily increases, see Fig.~\ref{fig:ts_normal}(b,e). These observations indicate that a continuous formation of new AC-domain nuclei initially, followed by gradual growth and merging. This means the AC-domains is able to overcome the invasion of D-domain and dominate the population.

Within M-areas, both the number of AC-domains and the average size of AC-domains show a consistent increase over time, and eventually stabilise as shown in right panels [Fig.~\ref{fig:ts_normal}(c,f)]. The results indicate the continual formation of new patches within M-areas, which also expand in size. However, unlike the scenario in AC-areas, these semi-stable AC-domains establish a competitive balance with D-domains during their growth. As a result, the isolated AC-domains are surrounded by D-domains and cannot merge with each other. This leads to the coexistence of two types of AC-domains along with D-domains.
\begin{figure*}[htbp!]
	\centering
	\includegraphics[width=0.9\textwidth]{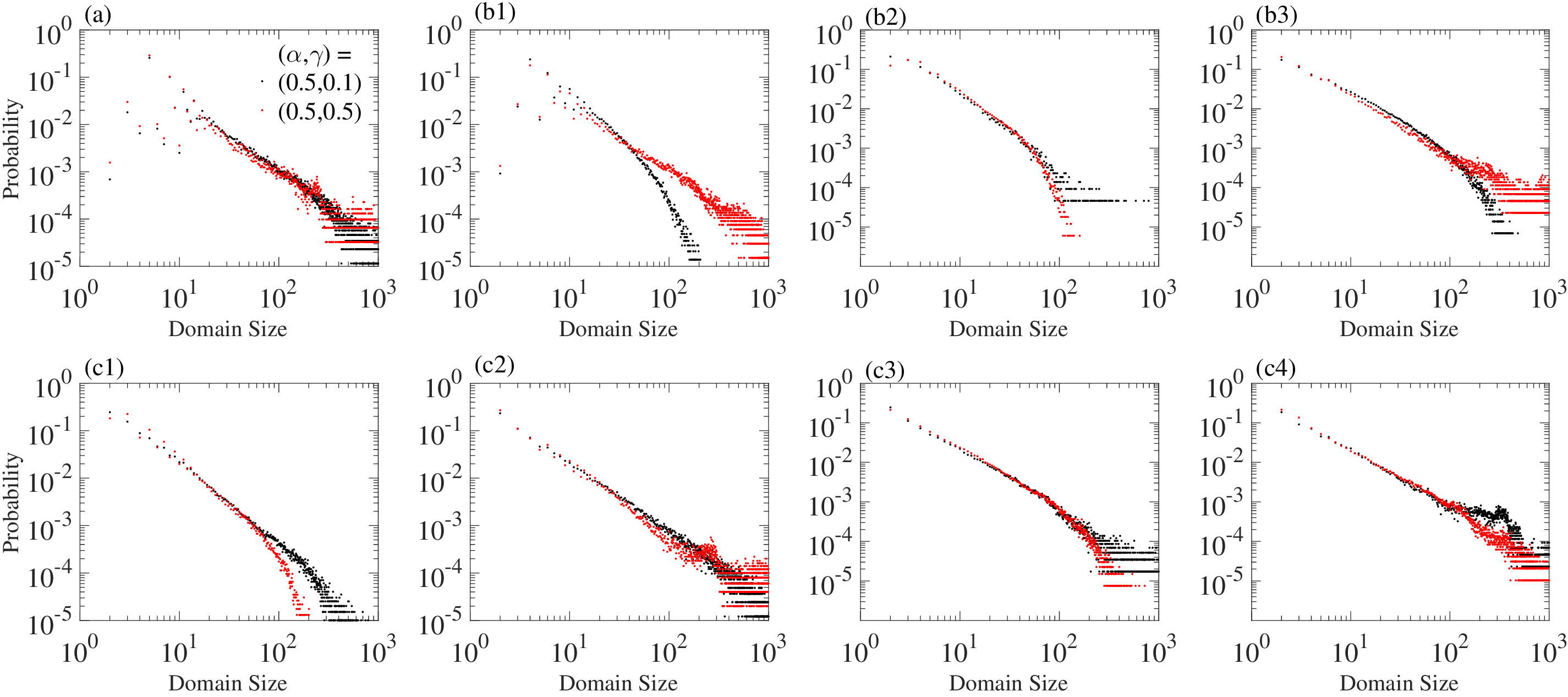}
	\caption{{\bf Distribution of primary anti-coordinated domain size around lower and upper boundaries of game parameter.} 
	(a), (b1-b3), and (c1-c4) respectively show the distributions of the size of AC-Domains on the von Neumann lattice, triangular lattice, and Moore lattice within double-logarithmic coordinates. The parameters of the panels are marked with the black circles in the corresponding panels of Fig.~\ref{fig:alpha_b}. 
	In (a), the game parameter $b = 0.143$, which is in close proximity to the lower boundary of game parameter $b_{l}^{*}$ for type-\RNum{1}.  
	In (b1-b3), $b = 0.09$, $0.58$ and $0.62$, which are proximate to $b_{l}^{*}$ and upper threshold $b_{u}^{*}$ for type-\RNum{1}, and $b_{l}^{*}$ for type-\RNum{2}, respectively. 
	In (c1-c4), $b = 0.305$, $0.334$, $0.82$ and $0.8665$, which are in close proximity to $b_{u}^{*}$ for type-\RNum{1}, $b_{l}^{*}$ and $b_{u}^{*}$ for type-\RNum{2}, and $b_{l}^{*}$ for type-\RNum{3}. 
	Other parameters: $\alpha = \gamma = 0.5$, $\epsilon =0.01$, and $|\mathcal{N}| = 10000$.
	} 
	\label{fig:ditribution_domain} 
\end{figure*}

Finally, we also provide the size distribution for primary AC-domains near $b_{l}^{*}$ and $b_{r}^{*}$ of different AC-areas in normal scenarios. These size distribution all follows either a power-law or an exponentially truncated power-law distribution, regardless of the lattice structure, see Fig.~\ref{fig:ditribution_domain}. Furthermore, the scaling exponent for a given $\alpha$ and the lattice type remains the same across different $\gamma$ for most cases. This observation may suggests that continuous phase transitions occur at $b_{l}^{*}$ and $b_{r}^{*}$~\cite{pascual2005criticality}, which is different from the discontinue phase transition in the previous works~\cite{wang2006memory}.

\subsubsection{Abnormal Scenarios}\label{subsec:abnormal_scenarios}

\begin{figure}[htbp!]
	\centering
	\includegraphics[width=0.75\textwidth]{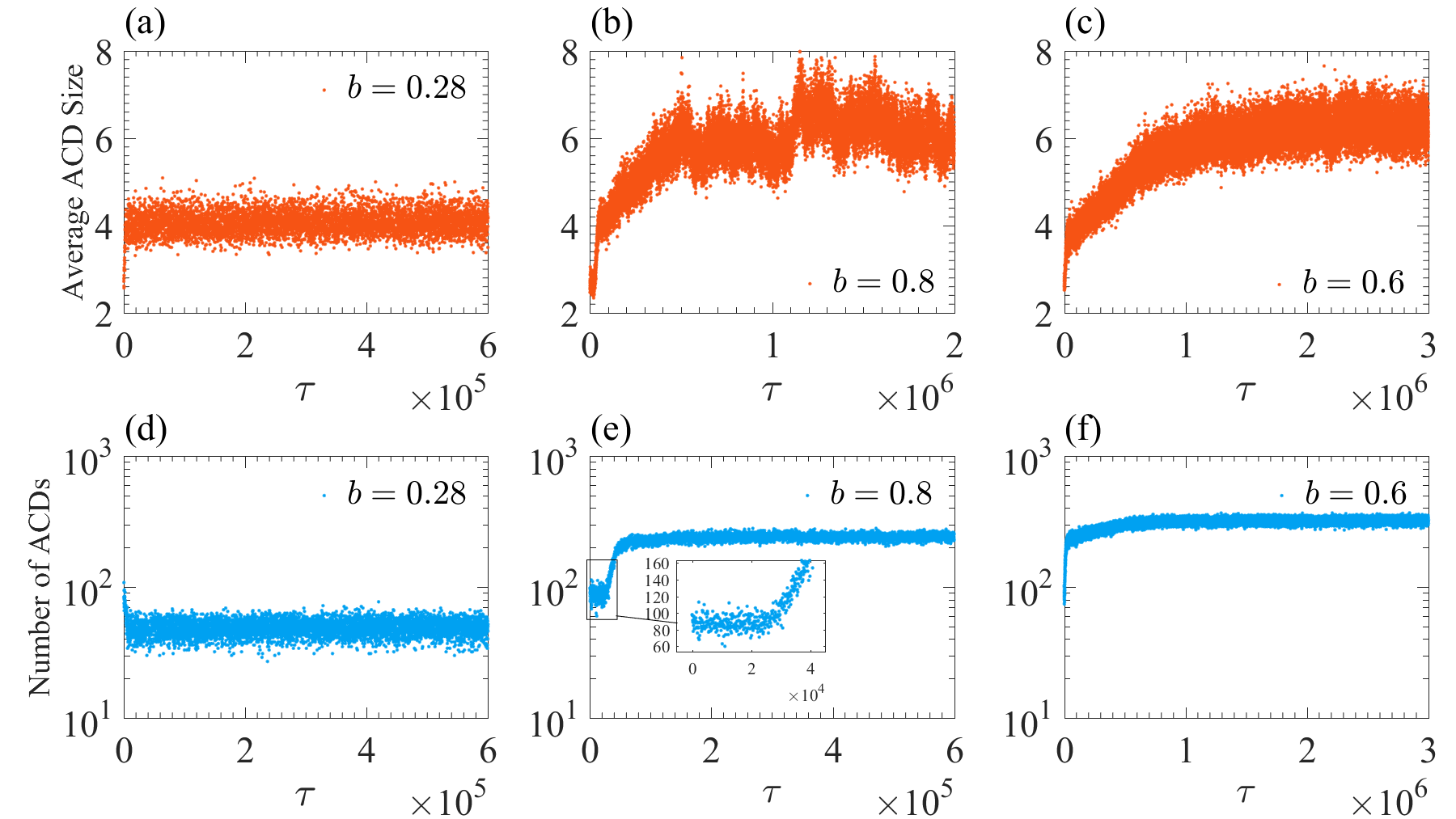}
	\caption{{\bf Statistics of anti-coordinated domains in the abnormal scenarios on a triangular lattice.} 
	We show the time series of the number of AC-domains and of the size of AC-domains within the NAC-area, the AC-areas, and M-areas respectively shown in the left, middle and right panels. 
	The learning parameters are $\tuple{\alpha, \gamma} = (0.9, 0.1)$, $ (0.1, 0.9)$ and $(0.5, 0.9)$ from left to right, respectively. 
	Parameters: $|\mathcal{N}| = 10000$.
	} 
	\label{fig:ts_abnormal} 
\end{figure}

When both the learning rate $\alpha$ and the discount factor $\gamma$ are large, or a large $\alpha$ combined with the frustrated lattice, abnormal scenarios are more likely to occur according to Figs.~\ref{fig:b} and~\ref{fig:alpha_b}.
To investigate these abnormal scenarios, we present the time series of the number of AC-domains and their average sizes in these scenarios, as shown in Fig.~\ref{fig:ts_abnormal}. 

For a large $\alpha$ on the frustrated lattice, the temporal evolution of both average size and number in AC-areas are similar to those in NAC-areas of normal scenarios, showing an increase in average size and a decrease in number, as Fig.~\ref{fig:ts_abnormal} (a,b) show. The results also reveal that the AC-domain nuclei exhibit instability, fail to maintain the domain growth.
This observation may explain the similarity in the trend of average cooperation preference with the increase of $b$ between these scenarios and the normal scenarios of NAC-areas. The reason why a large $\alpha$ leads to abnormal scenarios is that agents in this case tend to be more responsive to the actions of their neighbors. This increased sensitivity further amplifies the challenge of nucleation and growth of AC-domains, especially in the frustrated lattices. 

For the case where a large $\gamma$ is combined with large $b$, we respectively select the parameters in the theoretical AC-areas and M-areas as depicted in the middle panels [Fig.~\ref{fig:ts_abnormal} (b) and (e)] and right panels [(c) and (f)]. Similar to the normal scenarios of M-areas, both the number and average size are increased till stabilized. The observations demonstrate the nucleus of AC-domains are able to form and grow in the competition with D-domains. As a result, the AC-domains are unable to assert their dominance over the system. The results indicate a large $\gamma$ combined with a large $b$ may enhance the competitive advantage of D-domains. In fact, the abnormal scenarios being attributed to the fact that our assumptions regarding $1$-step recovery may not hold true, as the $\gamma$ determines the significance of future rewards. However, $1$-step recovery assumes agents to be short-sighted. 

\subsection{The compositions of population}\label{subsec:composition}
\begin{figure}[htbp!]
	\centering
	\includegraphics[width=0.85\textwidth]{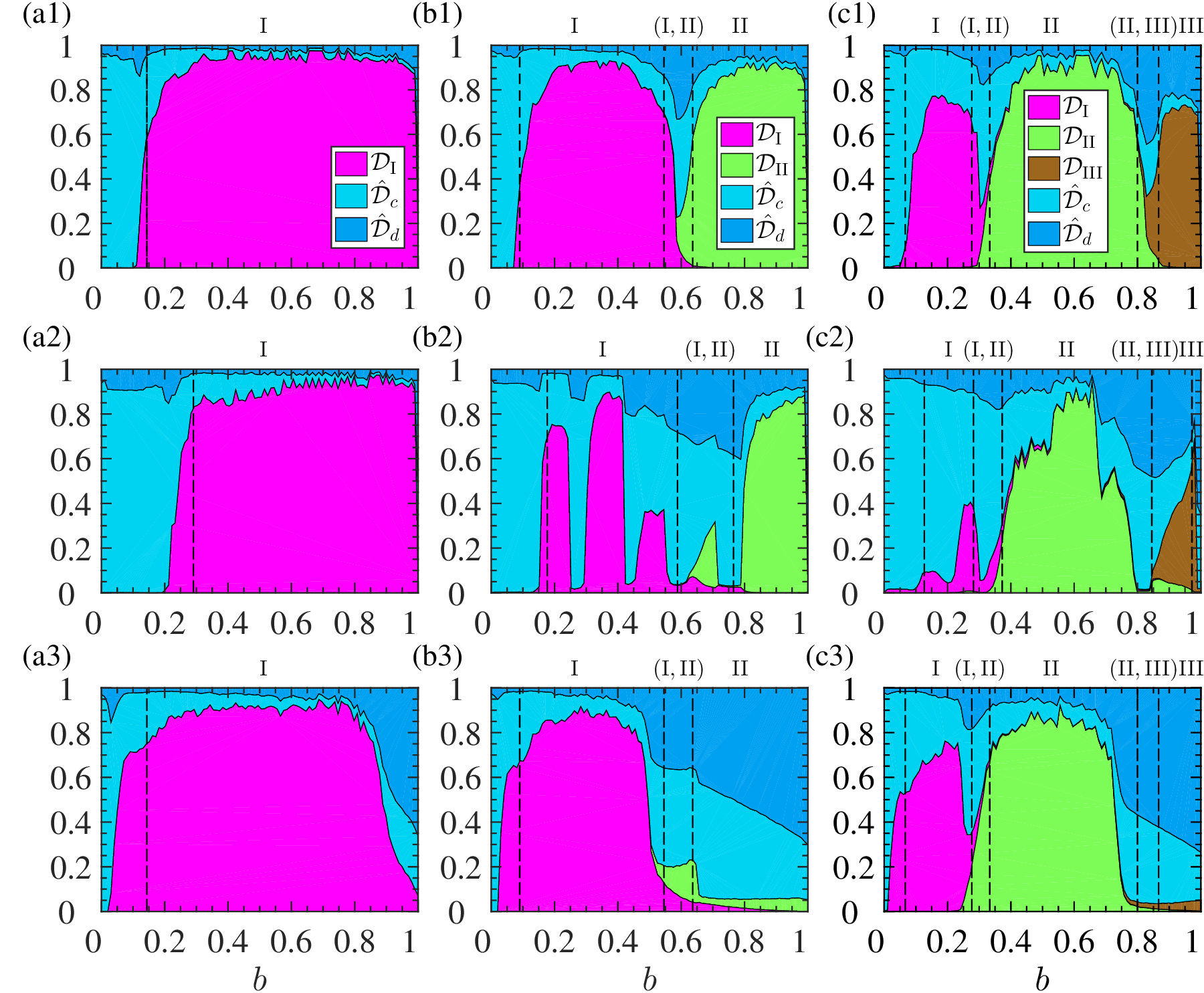}
	\caption{{\bf Compositions of AC-domains and cooperators/defectors in Non-Anti-coordinated domains.} 
	Here we distinguish different AC-domains as $\mathcal{D}_{X}$, where 	
	$X$ represents the type of the corresponding AC-domain. The individuals who cooperate and defect in the Non-AC domain are denoted as $\hat{\mathcal{D}}_{c}$ and $\hat{\mathcal{D}}_{d}$, respectively. Their proportions in the population correspond to the filling proportions in each panel.
	The combinations of learning rate and discounting rate, $\tuple{\alpha,\gamma}$, from top to bottom are  $\tuple{0.5,0.1}$, $\tuple{0.9,0.1}$ and $\tuple{0.5,0.9}$. 
	From left to right panel, the population topology corresponds to the von Neumann lattice, triangular lattice, and Moore lattice, respectively. 
	Other parameters: $\epsilon = 0.01$ and $|\mathcal{N}| = 10000$. 
	} 
	\label{fig:domain_composition} 
\end{figure}

As Eq.~(\ref{eq:fc_analysis}) shows, the average cooperation preference $\langle \bar{f}_{c}\rangle$ in population results from the compositions of different AC-domains and NonAC-domains. 
Here, we further provide detailed compositions of different AC-domains, and compositions of cooperators and defectors for NonAC-domains, see Fig.~\ref{fig:domain_composition}. 
The top panels in Figure~\ref{fig:domain_composition}(a1-c1) depict the normal scenarios that align with our theoretical predictions. These scenarios can be summarized as follows: 1) The AC-domains are nearly invisible in the NAC-area; 2) Each AC-area is dominated by a specific type-X AC-domains; 3) Within an M-area between type-$X$ and type-$X^{\prime}$, which act as transition area between AC-areas, two types of AC-domains coexist alongside NonAC-domains due to the semi-stability of both corresponding AC-domains. Besides, as the parameter $b$ increases in M-areas, one type-$X^{\prime}$ AC-domain gradually replaces type-$X$. 

The middle panels, especially Fig.~\ref{fig:domain_composition} (b2-c2), highlight the abnormal scenarios arising from large learning rates $\alpha$ in frustrated lattices. In such abnormal scenarios, the AC-domains fail to survive or flourish when confronted with the invasion of D-domains. Therefore, the proportion of any specific type of domain is very low within the population. In addition, the results suggest that frustrated lattices may be a necessary condition for the emergence of such abnormal scenarios, as no such scenario is observed in the von Neumann lattice. 

But for the combination of large $\gamma$ and large game parameters $b$, the abnormal scenarios appear to be irrelevant to whether the lattice is frustrated or not. This observation is highlighted in the bottom panels [Fig.~\ref{fig:domain_composition} (a3-c3)]. The occurrence of this phenomenon may be attributed to the violation of the assumption of $1$-step recovery, which is independent of the lattice as mentioned before.
Furthermore, the results also show that the cooperation preference in Non-AC domains exhibits a linear decrease as $b$ increases. 
\section{CONCLUSION AND DISCUSSION}\label{sec:conclusions}

In this study, we propose a model of Two-Agent Two-Action Reinforcement Learning Evolutionary Game ($2\times 2$ RLEG) and is applied to the snowdrift game (SDG) on three lattices. Interestingly, we observe the emergence of distinct and diverse anti-coordinated domains (AC-domains), which are expected but rarely seen in previous studies.
The emergence and types of these AC-domains depend on the learning parameters and the game parameter. Specifically, we can observe non-anti-coordination, anti-coordination, and mixed cases within the Non-Anti-Coordinated area (NAC-area), Anti-Coordinated areas (AC-areas), and Mixed areas (M-areas) of the parameter space, respectively. Furthermore, AC-areas can be further categorized based on their types of AC-domains. In the NAC-area, the occurrence of AC-domains is scarce. In contrast, within the AC-areas, only a specific type of AC-domain dominating. Whereas, within the M-areas located between AC-areas, we observe the coexistence of AC-domains accompanied by the presence of Disorder domains (D-domains).

To understand these observations, we developed a perturbation theory to investigate the stability of different types of AC-domains.
 Our theory successfully predicts different areas observed in simulations, and specifically we have the following findings: 1) Within the NAC-areas, no single type of AC-domain is stable. 2) In each AC-area, only a specific type of AC-domain exhibits stability. 3) However, within the M-areas located between AC-areas, we observe the presence of two semi-stable types of ACDs. Besides, in the ``\emph{normal scenarios}" conforming to the theory, our simulations further reveal there may be continuous phase transitions at the partitioning points between the areas~\cite{pascual2005criticality}, which is different from the discontinuous phase transition in the previous works~\cite{wang2006memory}.

Our study also reveals the specific ``\emph{abnormal scenarios}'', where the stable AC-domains are expected to dominate theoretically, but this is not the case or even fail to emerge in simulations.
 Based on the analysis of the nucleation and growth of AC-domains, we discover that agents with a large learning rate are more sensitive and responsive to the actions of their neighbors. This further makes the nucleation difficult, particularly with frustrated lattices. Moreover, the agents’ high expectation about the future deviates from our assumptions in theory and ultimately give rise to some abnormal scenarios.

Although our study provides new insights within the RLEG frame, there are still several open questions remained. 
Specifically, in order to fully understand the behavior of RLEG systems, it will be necessary to develop more sophisticated theories that incorporate the complex topologies of the real population~\cite{de1992isotropic,liu2019coevolution}. 
Furthermore, the computational complexity of RLEG is significantly higher compared to classical evolutionary games. This poses a challenge in identifying the type of phase transition and computing the associated critical exponents, especially conducting finite-size scaling~\cite{stanley1971phase,domb2000phase}. Therefore, improving the computational efficiency is crucial for simulating large-scale system of RLEGs. 
Finally, our work may act as a good starting point for studying spatiotemporal pattern emergence in RLEGs, but many important questions remain for future research.


\section*{Acknowledgments}
	We are supported by the Natural Science Foundation of China under Grants No. 12165014 and 12075144, and the Key Research and Development Program of Ningxia Province in China under Grant No. 2021BEB04032. 

\appendix
\section{The comparison between normal scenarios and abnormal scenarios}\label{sec:app_cmp}
\begin{figure}[htbp!]
	\centering
	\includegraphics[width=0.75\textwidth]{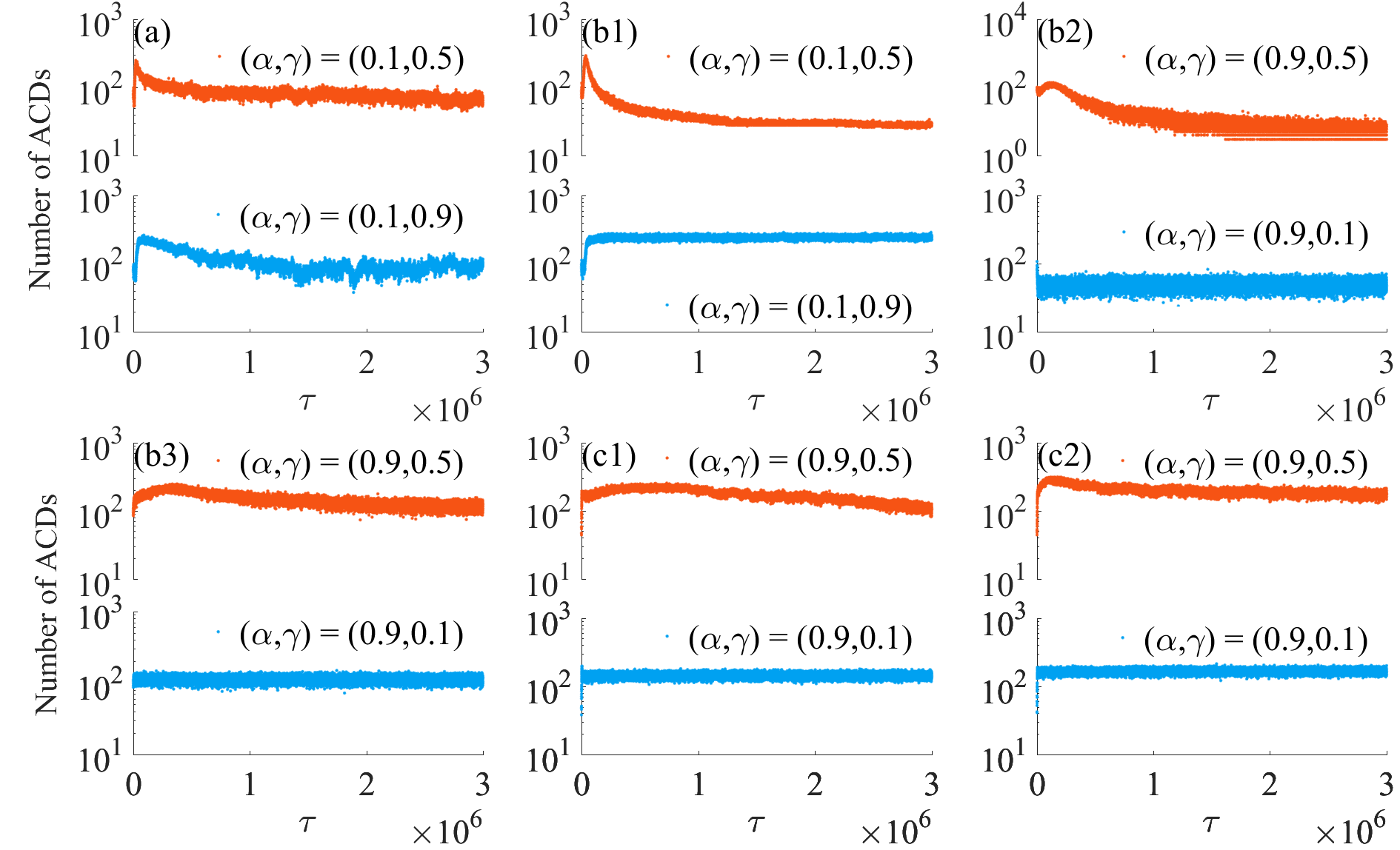}
	\caption{{\bf The comparisons of the time series of the number of AC-domains between normal scenarios and abnormal scenarios on different lattices.} In Panel (a), the comparison is made on von Neumann lattices with a game parameter $b = 0.95$. In Panels (b1-b3), the comparisons are on a triangular lattice with corresponding $b = 0.8$, $0.28$ and $0.75$. In Panels (c1-c2), the comparisons are on a Moore lattice with corresponding $b = 0.2$, $0.32$. In each panel, the upper part corresponds to the results in normal scenarios, while the lower part represents the results in abnormal scenarios.
		The sharing learning parameters in all panels are exploration rate: $\epsilon = 0.01$ and the scale of the population: $|\mathcal{N}| = 10000$.} 
	\label{fig:app_ts_num} 
\end{figure}

\begin{figure}[htbp!]
	\centering
	\includegraphics[width=0.75\textwidth]{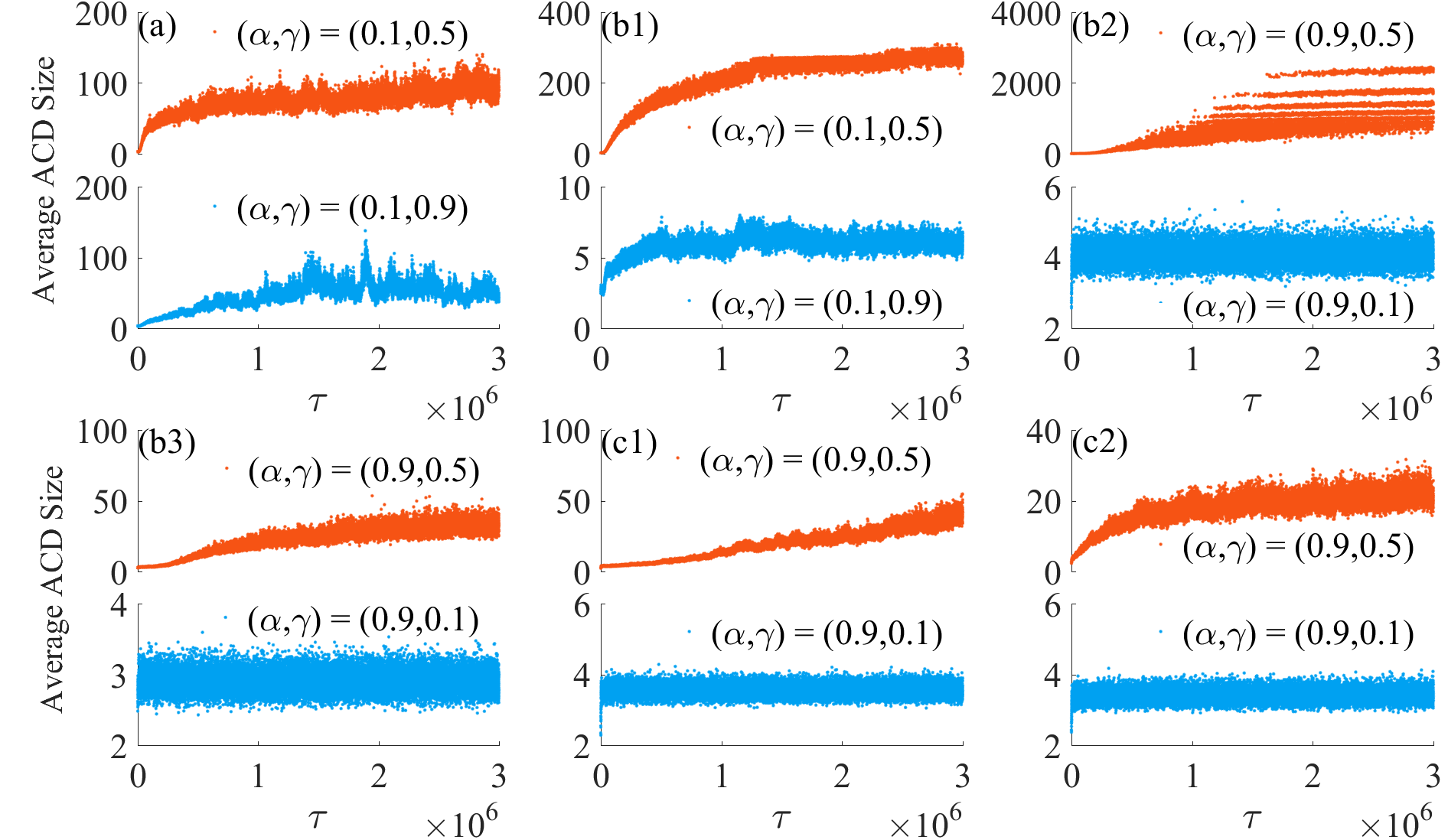}
	\caption{{\bf The comparisons of the time series of the average size of AC-domains between normal scenarios and abnormal scenarios on different lattices.} The parameters in the panels are identical to the corresponding panels in Fig.~\ref{fig:app_ts_num}. In each panel, the upper part corresponds to the results in normal scenarios, while the lower part represents the results in abnormal scenarios.}\label{fig:app_ts_ave_size} 
\end{figure}

In this section, we compare the time series of the number and average size of Anti-Coordinated Domains (AC-domains) on different regular lattices between normal and abnormal scenarios, as shown in Figs.~\ref{fig:app_ts_num} and~\ref{fig:app_ts_ave_size}. In normal scenarios, regardless of the type of lattice, the panels show an initial increase followed by a decrease in the number of AC-domains over time. However, the average size of AC-domains consistently increases and remains unchanged at the end. These results confirm our analysis in Fig.~\ref{fig:ts_normal} of Sec.~\ref{subsec:normal_scenarios} that AC-domains undergo three processes: nucleation, growth, and eventual merger.

In abnormal scenarios where agents exhibit a high learning rate $\alpha$ on the frustrated lattice, both the number and average size of AC-domains remain unchanged over time [referring to (b2-b3) and (c1-c2)]. These results reinforce the findings in Sec.~\ref{subsec:abnormal_scenarios} that randomly generated nuclei of AC-domains lack stability. Agents with a high learning rate are overly sensitive to their neighbors’ actions, compromising the stability of their Q-table structure.

In another abnormal scenario where agents have high expectations for the future ($\gamma$), the number of AC-domains initially increases followed by a small decrease or remaining unchanged, while the average size of AC-domains consistently increases and remains unchanged at the end. The presence of high expectations for the future may lead to decreased stability of AC-domains when faced with the invasion of Disordered domains (D-domains). However, over time, a balance is established to counteract the invasion and maintain stability. The abnormal scenarios deviating theory is due to the invalid of the assumption in our theory.

\begin{figure}[htbp!]
	\centering
	\includegraphics[width=0.75\textwidth]{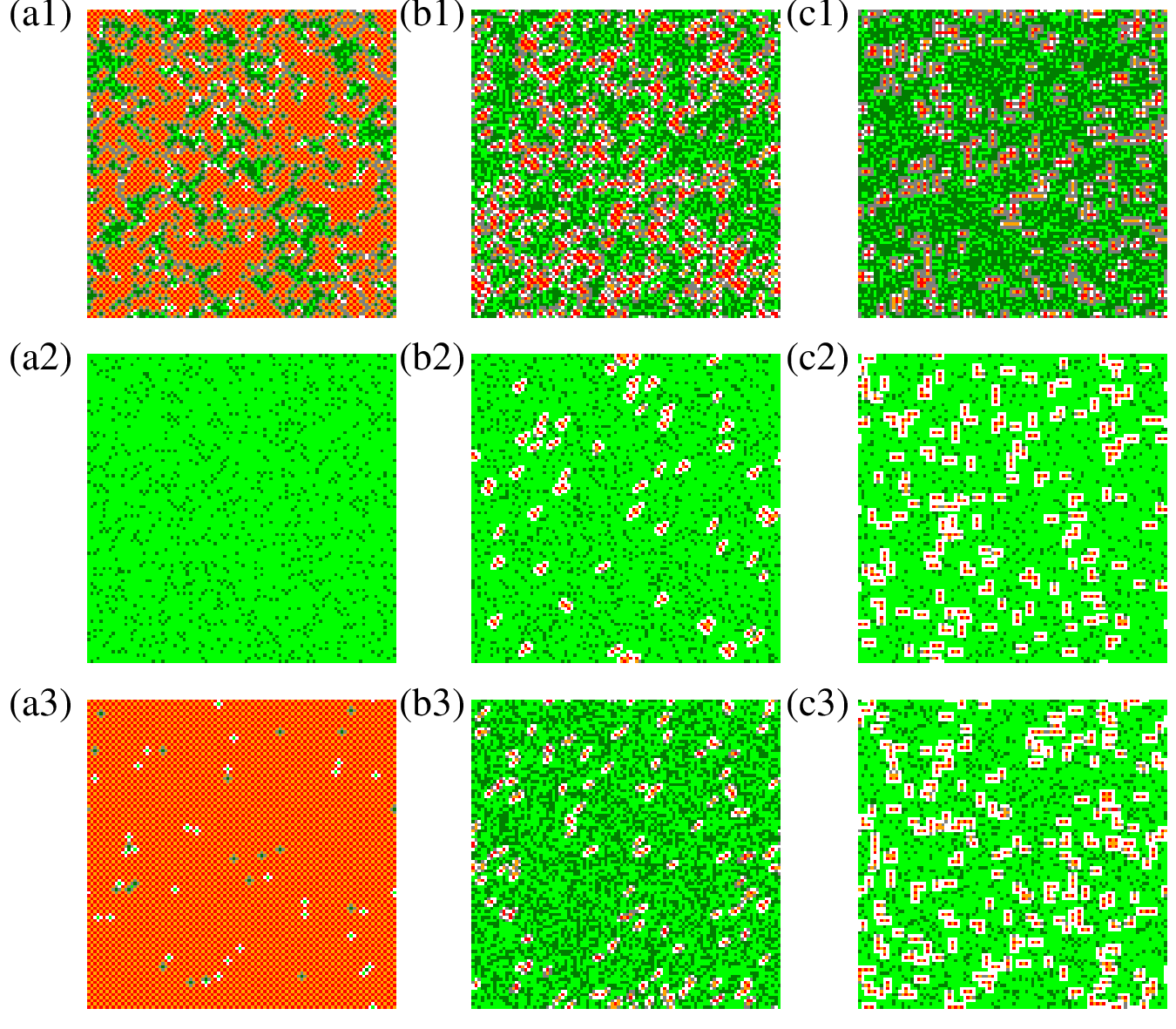}
	\caption{{\bf The final snapshots of abnormal scenarios on different lattices.} In panels (a1-a3), the snapshots show the agent’s actions on von Neumann lattices. The corresponding panel parameters are $\tuple{\alpha,\gamma, b} = \tuple{0.1, 0.9, 0.95}$, $\tuple{0.9, 0.1, 0.1}$, and $\tuple{0.9, 0.1, 0.75}$.
		In panels (b1-b3), the snapshots show the agent’s actions on triangular lattices. The corresponding panel parameters are $\tuple{\alpha,\gamma, b} = \tuple{0.1, 0.9, 0.8}$, $\tuple{0.9, 0.1, 0.28}$ and $\tuple{0.9, 0.1, 0.75}$.	In panels (c1-c3), the snapshots show the agent’s actions on Moore lattices. The corresponding panel parameters are $\tuple{\alpha,\gamma, b} = \tuple{0.1, 0.9, 0.9}$, $\tuple{0.9, 0.1, 0.2}$ and $\tuple{0.9, 0.1, 0.32}$.		
		In the panels, the cooperators and defectors are marked by \domainc{} and \domaind{}, respectively, within the AC-domains, by \wallc{} and \walld{} on the walls, and by \disorderc{} and \disorderd{} within the D-domains. The sharing learning parameter in (a-c)  are exploration rate: $\epsilon = 0.01$ and the scale of the population: $|\mathcal{N}| = 10000$.} 
	\label{fig:snapshots2} 
\end{figure}

In Fig.~\ref{fig:snapshots2}, we present additional snapshots of abnormal scenarios on different lattices after the system has reached stability. As depicted in the top panels (a1-c1), the coexistence of AC-domains and D-domains is observed in the abnormal scenarios caused by the high future expectations of agents. Besides, we further notice that the average size of AC-domains on the von Neumann lattice is larger compared to the triangular and Moore lattices. This finding reinforces the notion of a competitive balance between AC-domains and D-domains, but it also suggests that the frustration effect might contribute to the increased instability of AC-domains. 

In abnormal scenarios resulting from the high learning rate of agents, we do not observe the actual patterns of AC-domains. Instead, we only see randomly generated nuclei of AC-domains in the snapshots, as depicted in panels (a2-c2) and (b3-c3). This indicates that the nuclei of AC-domains are unstable and cannot grow into their intended patterns in such scenarios.

\section{The robustness of anti-coordinated pattern}\label{sec:app_robust}
\begin{figure}[htbp!]
	\centering
	\includegraphics[width=0.75\textwidth]{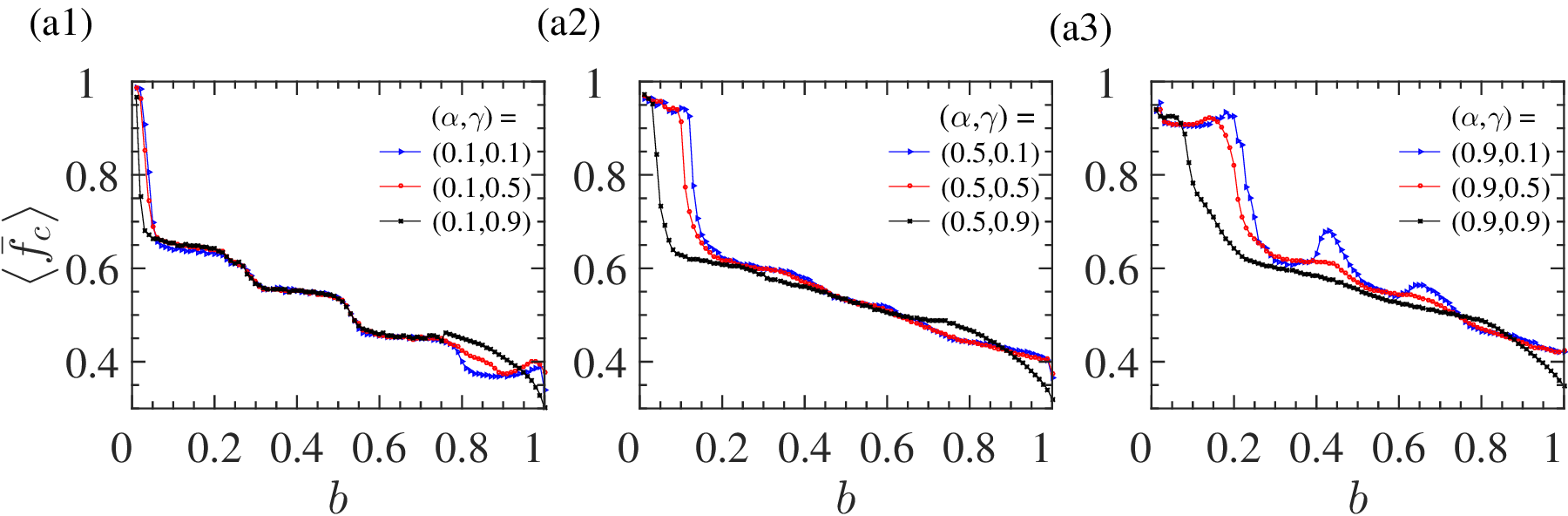}
	\caption{{\bf The comparisons of the time series of the average size of AC-domains between normal scenarios and abnormal scenarios on different lattices.} The parameters in the panels are identical to the corresponding panels in Fig.~\ref{fig:app_ts_num}. In each panel, the upper part corresponds to the results in normal scenarios, while the lower part represents the results in abnormal scenarios.}\label{fig:app_rr} 
\end{figure} 
Here, we further investigate the robustness of the anti-coordination pattern across different structured populations. For comparison with the von Neumann lattice, we assign agents to regular random networks~\cite{newman2002random,erdHos1960evolution}, ensuring each agent has the same number of neighbors as in the von Neumann lattice. As shown in Figure~\ref{fig:app_rr} (a1), we observe that the average cooperation preference, as a function of game parameter, exhibits some ``\emph{flat areas}'' when the agents have a low learning rate. Based on the analysis in the main text, we can reasonably conclude that these flat areas result from the anti-coordinated effects, indicating the robustness of the anti-coordination phenomenon across different populations. However, unlike the regular lattice, these flat areas disappear as the learning rate equipped by agents increases [see (a2) in Fig.~\ref{fig:app_rr} and (a2) in Fig.~\ref{fig:b}]. This suggests that the anti-coordination phenomenon is more vulnerable to the learning rate on regular random networks. 

In addition, we also observed some abnormal scenarios similar to those described in the main text. Specifically, we found that a high value of $\alpha$ leads to increased volatility of the flat areas as game parameter $b$ changes, given a low expectation of future $\gamma$. Furthermore, a combination of high $\gamma$ and high game parameter 
$b$ can exacerbate the deterioration of the anti-coordination phenomenon [see Fig.~\ref{fig:app_rr} (a1) and (a3)]. The results demonstrate that phenomena observed on the lattice may be general in structured population.

\bibliographystyle{elsarticle-num-names} 
\bibliography{main}

\end{document}